\documentclass[11pt]{article}
\usepackage{todonotes}

\newcommand{\bhedit}[1]{{\color{black} #1}}

\usepackage[top = 3cm, bottom =3.5cm, right = 2.5cm, left = 2.5cm]{geometry}

\usepackage{amsmath,
  enumerate,amsfonts,subcaption}
\usepackage{graphicx}
\usepackage{overpic}

\newcommand{\mcl}[1]{\mathcal{#1}}

\newcommand{\reals}{\mathbb{R}}
\newcommand{\complex}{\mathbb{C}}

\newcommand{\mb}[1]{\mathbf{#1}}

\newcommand{\diag}{\text{diag}}

\newcommand{\email}[1]{\texttt{#1}}

\title{{A Bayesian approach for energy-based estimation of acoustic aberrations in 
high intensity focused ultrasound treatment}\thanks{This work was
supported in part by the Natural Sciences and Engineering
Research Council of Canada,
the Brain Canada Multi-Investigator
Research Initiative and the Focused Ultrasound Foundation.}}

\author{
Bamdad Hosseini \thanks{Corresponding author.} \thanks{Department of Computing and Mathematical 
Sciences, California Institute of Technology, Pasadena, CA, 91125, USA (\email{bamdadh@caltech.edu}).} 
   \and 
Charles Mougenot \thanks{University Medical Center Utrecht, Heidelberglaan 100, 3584 CX Utrecht, Netherlands  (\email{c.mougenot@umcutrecht.nl}).} \and 
Samuel Pichardo \thanks{Radiology and Clinical Neurosciences, School of Medicine, University of Calgary, AB, T2N 1N4, Canada
    (\email{samuel.pichardo@ucalgary.ca}).} \thanks{Hotchkiss Brain Institute, University of Calgary, AB, T2N 1N4, Canada.} \and 
Elodie Constanciel \thanks{Hospital for Sick Children, 555 University Avenue,
  Toronto, ON, M5G 1X8, Canada
  (\email{elodie.constanciel@sickkids.ca}, \email{james.drake@sickkids.ca}).} \and 
James M. Drake \footnotemark[7] \and
John M. Stockie \thanks{Department of Mathematics, Simon Fraser University, Burnaby, BC, V5A 1S6, Canada (\email{stockie@math.sfu.ca}). }
}

\begin{document}



\maketitle

\begin{abstract} 
High intensity focused ultrasound is a non-invasive method for
treatment of diseased tissue that uses
a beam of ultrasound 
to generate heat within a small volume.
A common challenge in
application of this technique is that heterogeneity of the biological
medium
can defocus the ultrasound beam.
Here we reduce the problem of refocusing the beam  to the 
 inverse problem of estimating the acoustic aberration due to
the biological tissue
from acoustic radiative force imaging data.
We solve this inverse problem using a Bayesian framework with a 
hierarchical prior and solve the inverse problem using a
Metropolis-within-Gibbs algorithm. The framework is tested using
both synthetic and experimental datasets.
We demonstrate that our approach has the ability to estimate the aberrations 
using  small datasets, as little as 32 sonication tests, which can lead to significant speedup in 
the treatment process. Furthermore, our approach is compatible
 with
a wide range of sonication tests  and can be applied to other
energy-based measurement techniques.
\end{abstract}
\vspace{2pc}
\vspace{2pc}

\newpage


\section{Introduction}\label{sec:introduction}
High intensity focused ultrasound (HIFU) treatment is a non-invasive method for treatment of 
diseased tissue. The treatment uses a focused beam of ultrasound waves that converge 
onto a focal point. The resulting absorption of ultrasound
generates heat which in turn can ablate the
targeted tissue.
The method 
has shown clinical success in treatment of uterine fibroids
\cite{fennessy, ikink, stewart-clinical}, prostate
cancer \cite{crouzet}, liver tumours \cite{illing-safety,wu2004}, brain
disorders \cite{elias2013pilot,jeanmonod,lipsman} and 
other medical conditions \cite{liberman}. However,
application of this method for
treatment of brain tissue remains a challenge.
 Strong aberrations due to the skull
bone, specifically the shift in the phase of the acoustic signal, defocus the beam 
and result in a loss of acoustic pressure.
This problem can be resolved by estimating the introduced
acoustic aberrations. If the estimate is accurate enough then one can
compensate the phase of the acoustic signals (at the transducer) and refocus the
beam behind the skull bone. 

One approach for estimating the aberrations is to use Magnetic
Resonance (MR) imaging \cite{hynynen} or Computed Tomography (CT)
\cite{aubry-CT, marquet-CT}  to obtain a three
dimensional model of the patient's skull and use this information in a
computer model for acoustic wave propagation to
 estimate the tissue aberration and the  phase
shift needed  to refocus the beam. However, this approach
is limited by both the computational cost of the
model and the accuracy of the estimates for the  properties of the tissues.

An alternative approach is the
so called energy-based focusing techniques of
\cite{herbert-energy, larrat}. Here, Magnetic Resonance
Acoustic Radiation Force Imaging (MR-ARFI) is used to obtain
measurements of the intensity of
the acoustic field at the focal point. 
MR-ARFI uses low-duty cycle HIFU pulses that generate tissue
displacement in the order of microns at the focal point of the beam. The small displacement is measured with MRI using gradient pulses that encode the tissue displacement in the phase information of an MR image 
\cite{chen2010optimization,maier}. Using ARFI, displacement maps are generated and can be used to verify and correct the degree of focusing of HIFU beam 
\cite{marsac2012mr}.
The energy-based focusing techniques in 
\cite{herbert-energy, larrat} 
use a dataset of displacement maps that is generated by imposing
specific excitation patterns at the ultrasound transducer. Columns
of a Hadamard
matrix are used  in
 \cite{herbert-energy} while \cite{kaye} uses  Zernink polynomials.
Afterwards,
the resulting displacement maps are used to estimate both the acoustic
field of the transducer and the aberrations induced by the ultrasound
propagation medium. The main drawback of this
technique is the need for a large number of sonication
tests which requires a long acquisition time for the MRI data.
 Recently, it was argued in \cite{liu-hifu}
that energy-based techniques can be cast as a 
penalized least-squares problem which enables one to use more
general
excitation patterns. They showed that using randomized calibration
sequences can reduce the number of sonication tests significantly.

{
In this article, a far-field approximation to the 
three dimensional acoustic equation is used as a forward
model that can be evaluated efficiently. The
effect of the tissue is modelled as an infinitely thin
aberrator in front of the transducer, following
\cite{liu-hifu}. These assumptions allow us to use a fast forward map that 
can be evaluated many times for the purposes of estimating the aberrations and 
quantifying uncertainties. Furthermore, 
the estimation of aberrations can 
be written in the form of a phase retrieval problem where the function 
is typically recovered from the amplitude of its Fourier transform
\cite{fienup2013phase, marchesini2007invited, marchesini2007phase, 
  shechtman2015phase}; thus, the methodology of this
article is also applicable to 
 phase retrieval problems in other applications. 
}

The goal of this article is to demonstrate the feasibility of using a new Bayesian method to
estimate the acoustic aberrations with a small number of
sonication tests. The central idea is to cast the problem within the framework
of Bayesian inverse problems.
The Bayesian perspective provides a general framework for estimation
of parameters that model the aberrations
from a finite set of measurements. Appropriate models are
chosen to explain the data, the measurement noise and prior knowledge
of the parameters. Afterwards, 
an entire probability distribution on
the parameters is obtained rather than a single point
estimator. The Bayesian formulation
can also be viewed as a generalization of the least-squares
formulation of \cite{liu-hifu} (minimizers of penalized least-squares
functionals are often equivalent to maximizers of the density of
an underlying posterior distribution when the parameters are finite dimensional \cite{somersalo}). This allows for stable estimation of the
aberrations with very noisy data and few sonication
tests.

Furthermore, the Bayesian approach provides an estimate of the
parameters as well as the associated uncertainties in that estimate.
An introduction to the Bayesian perspective for solution of
inverse problems 
and many of the techniques that are used in this article
can be found in the monographs \cite{calvetti, somersalo, tarantola} and
the article \cite{stuart-acta-numerica} as well as the
references therein. The
Bayesian approach to inverse problems has been successfully applied in various areas of medical imaging
such as electrical impedance tomography \cite{kaipio-EIT}, optical
diffuse tomography \cite{arridge-dot}
and dynamic X-ray tomography \cite{lassas-x-ray} as well as other
fields such as astronomy \cite{edwards-supernovae, pursiainen} 
and geoscience \cite{fox-geothermal, Iglesias-subsurface, tarantola}.

Our Bayesian approach consists of 
 constructing a hierarchical smoothness prior for the aberration
 that reflects the prior knowledge that the aberration
parameters tend to change smoothly between nearby elements on the
transducer; that is, the properties of the tissue do not change
dramatically between elements.
 Combining this prior knowledge with the forward model and
the data, results in a posterior distribution which is viewed as the
solution to the inverse problem. A Metropolis-within-Gibbs (MwG) sampler
\cite{liu-mc,casella} is
used for exploring this distribution and obtaining several
statistics such as the posterior mean and standard deviation of the
aberrations as well as independent samples. 

The remainder of this article is organized as follows.
Section~\ref{sec:mathbackground} is dedicated to the mathematical theory and
the setup of
the problem. A brief introduction to the far-field approximation of the
acoustic equation is presented which is followed by the setup of the 
forward model that explains the MR-ARFI data. Next, the formulation of the
Bayesian inverse problem is discussed where the likelihood and prior
distributions are constructed. At the end of this section a MwG algorithm
for sampling the posterior distribution is proposed. 
Section~\ref{sec:methods}
concerns the setup of the test for synthetic and experimental conditions
that
were performed to verify  the method.
 The results are then presented in
Section~\ref{sec:results} which is followed by a discussion of the mathematical
framework and the results in Section~\ref{sec:discussion}. 
\begin{figure}[htp]
  \centering
  \begin{subfigure}{0.4 \textwidth}
  {\includegraphics[width=.9\textwidth, clip=true, trim=0cm 0cm 0cm
  0cm, bb = 0 0 360 360]{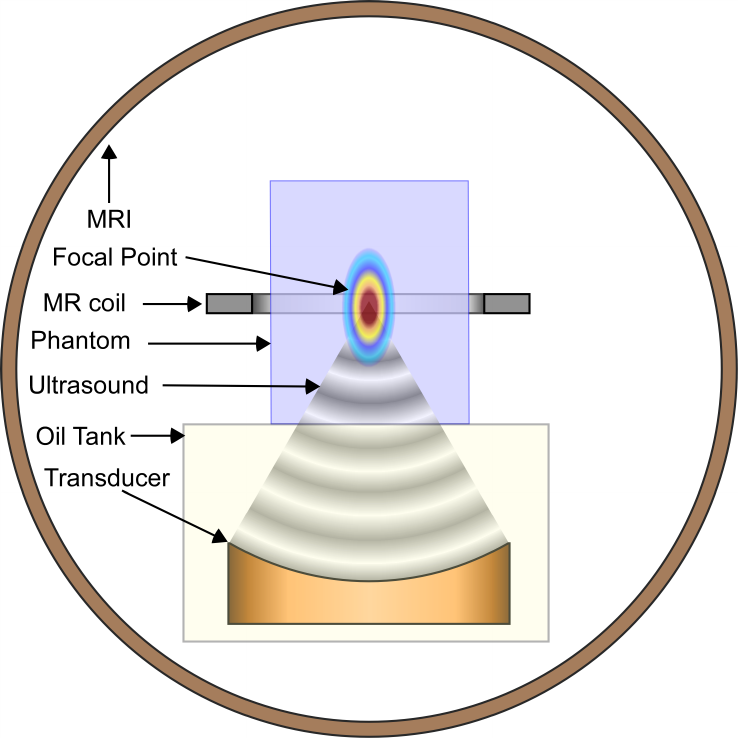} }
\caption{}  
  \end{subfigure}
  \begin{subfigure}{0.4 \textwidth}
  \raisebox{1.cm}{\includegraphics[width=1 \textwidth, clip=true, trim=1cm 6cm 
30cm
  10cm]{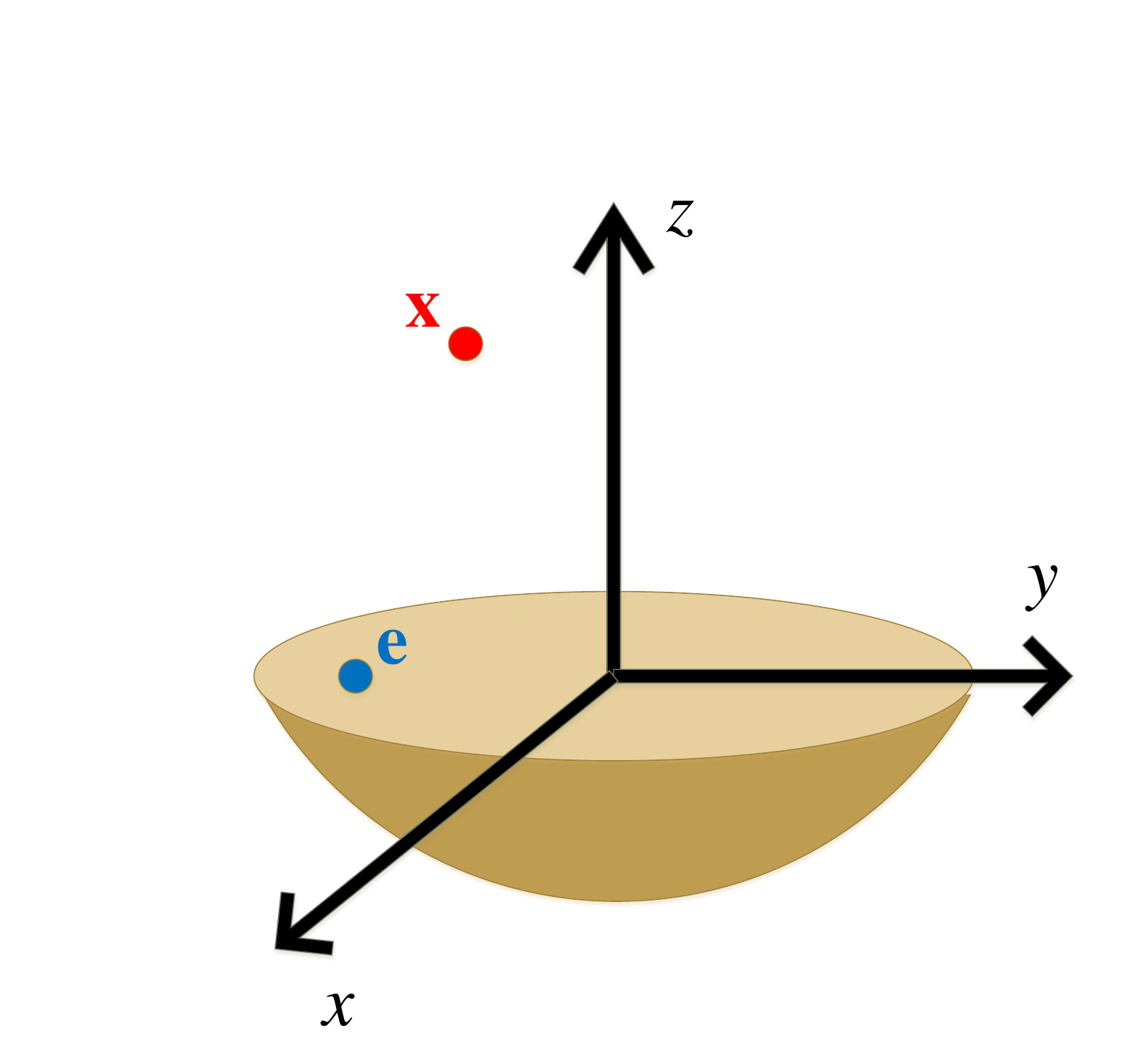}}
\caption{}  
  \end{subfigure}
  \caption{(a) Schematics of the high intensity focused ultrasound
    setup. The transducer is submerged in an oil tank and generates an
    acoustic field which is directed at a phantom. The resulting
    displacements are measured using an MR-coil inside an MRI machine.
    (b) The coordinate system of the transducer
that is used in equations
    \eqref{plane-wave}--\eqref{pressure-model}
for computing the acoustic field.} 
  \label{fig:hifu-schematic}
\end{figure}

\subsection{Notation}
Throughout this article lower case and Greek bold letters
denote vectors and upper case bold letters denote
matrices. Given a matrix $\mb{A}$ we use $\mb{A}^\ast$ to denote its
adjoint. Finally, given $\mb{m} \in \reals^N$ and a positive
definite matrix $\pmb{\Sigma} \in
\reals^{N\times N}$,  we use $\mcl{N}(\mb{m}, \pmb{\Sigma})$ to denote a
Gaussian random variable with mean $\mb{m}$ and covariance matrix $\pmb{\Sigma}$.
\section{Mathematical background and theory}\label{sec:mathbackground}
\subsection{Forward problem}\label{section:forward-problem}
A far-field approximation to the acoustic field of
a single piezoelectric ultrasound emitter is obtained under the assumption that 
each ultrasound emitter is infinitesimally small and emits a 
radially symmetric acoustic wave of amplitude $p_0$
[Pa] and frequency
$\omega$ [Hz]. Then the pressure field $p(\mb{x})$ [Pa] at a location $\mb{x}$ [m],
generated by a piezoelectric emitter at location $\mb{e}$ [m]
(see Figure \ref{fig:hifu-schematic}) is given by
\begin{equation}\label{plane-wave}
 p(\mb{x}; \mb{e}) = zf(\mb{x}; \mb{e}) \quad \text{where} \quad  z= p_0 \exp( i \omega
t) \quad \text{and} \quad f(\mb{x};\mb{e})= \frac{1}{|\mb{x}-\mb{e}|}\exp\left( i \frac{\omega}{c_0}
|\mb{x} -
\mb{e}|\right).
\end{equation}  
Here
$c_0$ [m/s] is the speed of
sound and $f(\mb{x}; \mb{e})$ is referred to as the free-field of the
emitter \cite{liu-hifu}.
Interaction of the acoustic waves with the
tissue results in dissipation of 
energy  and a shift in the phase as a result of friction (micro-scale displacements) and
scattering \cite{Haar-HIFU}.
The acoustic pressure $\tilde{p}(\mb{x})$ in the presence of the tissue
can be modelled as
\begin{equation} \label{lossy-field}
\tilde{p}(\mb{x}; \mb{e}) = \mu p(\mb{x}; \mb{e})\quad
\text{where} \quad \mu = \zeta \exp(i \phi t).
\end{equation}
Here, $\zeta$ is the amplitude coefficient related to the attenuation of the medium  and $\phi$ is the shift in phase.
 
A physical transducer can be a phase array consisting of a large number of  piezoelectric emitters.
Let $N$ denote the number of physical piezoelectric emitters ($N=256$ in the
case 
of the Philips Sonalleve V1 system (Philips, Healthcare, Vantaa, Finland)
that is used in the experiments of Section~\ref{methods:physical-test}) and
let $\mb{z}\in \complex^N$ denote
 the sonication pattern at the transducer which is the
vector containing the phase and amplitude of the acoustic waves transmitted by each emitter. Also, $\mb{a} \in
\complex^N$ is defined as the vector of aberrations pertaining to each element, \bhedit{ obtained 
by concatenating the $\mu$
variable of \eqref{lossy-field} for all elements on the transducer}. Finally, define the field of view to
be the displacement data of a
region of $\sqrt{M}\times \sqrt{M}$
voxels
around the focal point which is extracted from MR-ARFI
images ($\sqrt{M} = 19$ in Section~\ref{sec:results} for the synthetic test
and $\sqrt{M}=7$ for the physical experiment). 
Let $\tilde{\mb{p}} \in \complex^{M}$ denote the 
measured values of the pressure at each voxel in presence of
aberrations concatenated into a long vector. 
Then
\begin{equation}
  \label{pressure-model}
  \tilde{\mb{p}} = \tilde{\mb{F}} \: \diag(\tilde{\mb{z}}) \mb{a},
\end{equation}
where $\diag(\tilde{\mb{z}})$ is the diagonal matrix created from the
entries of $\tilde{\mb{z}}$ and $\tilde{\mb{F}} \in \complex^{M
  \times N}$ is referred to as
the free-field matrix \cite{liu-hifu} which is the mapping of
the pressure in the absence of aberrations. { The $i,j$ entry of 
$\tilde{\mb{F}}$ is given by the value of the free-field function $f(\mb{x}_i; \mb{e}_j)$ in \eqref{plane-wave}
of a piezoelectric element at location $\mb{e}_i$ evaluated at a 
 voxel centered at $\mb{x}_j$, in the absence of the aberrator. Thus, $\tilde{\mb{F}}$ depends solely on the geometry of the 
transducer and the location of the voxels, and $\tilde{\mb{z}}$ are 
specified by the user through the design of the sonication patterns and so 
 both $\tilde{\mb{F}}$ and $\tilde{\mb{z}}$ are known a priori.}

   The above model for the pressure field $\tilde{\mb{p}}$ relies on the assumption
that tissue aberrations 
 can be modelled 
as an infinitesimally thin aberrator in front of the transducer
\cite{liu-hifu}. The measured displacement in MR-ARFI images is proportional to the 
total intensity of the signal which is equal to the square of the
modulus of pressure \cite{maier}. The constant of proportionality is
generally unknown but it can be estimated in a calibration step as 
discussed in Section~\ref{sec:calibration}. Throughout the remainder of this
article the constant of proportionality is accounted for in the free-field
matrix.
To
this end, let
$\tilde{\mb{d}} \in \reals^M$ be the vector of displacements at
each voxel. Then 
\begin{equation}
  \label{displacement-model}
  \tilde{\mb{d}} = \diag(\tilde{\mb{p}}){\tilde{\mb{p}}^\ast}
\end{equation}
where $\tilde{\mb{p}}^\ast$ denotes the element wise complex
conjugate of $\tilde{\mb{p}}$. In practice, a finite number of $J$
sonication tests are performed where vectors
$\tilde{\mb{z}}_j$ for $j =1,\cdots, J$ are prescribed as input at the transducer and give
rise to MR-ARFI images. Each image can be summarized as a vector of displacements
$\tilde{\mb{d}}_j$. These measurements constitute a dataset that
is used to estimate the aberrations $\mb{a}$. Then, a model is needed 
in order to relate $\mb{a}$ to the entire displacement dataset. Define
the matrices
\begin{equation}\label{transducer-formulation}
\begin{aligned}
&\mb{Z} :=  
\begin{bmatrix}
  \diag(\tilde{\mb{z}}_1) \\ 
  \diag(\tilde{\mb{z}}_2) \\ 
  \vdots \\
    \diag(\tilde{\mb{z}}_J)
\end{bmatrix} \in \complex^{JN \times N}, \quad
&& \mb{F} := \mb{I}_{J\times J} \otimes \tilde{\mb{F}} \in 
\complex^{JM \times JN}, \\ 
& \mb{p} :=
\begin{bmatrix}
  \tilde{\mb{p}}_1 \\
\tilde{\mb{p}}_2 \\
\vdots \\
\tilde{\mb{p}}_J
\end{bmatrix} \in \complex^{JM}, \quad 
&&\mb{d} :=
\begin{bmatrix}
  \tilde{\mb{d}}_1 \\
\tilde{\mb{d}}_2 \\
\vdots \\
\tilde{\mb{d}}_J
\end{bmatrix} \in \reals^{JM},
\end{aligned}
\end{equation}
where $\mb{I}_{J\times J}$ is the $J\times J$ identity matrix and
$\otimes$ denotes the  Kronecker product. Then the forward model
can be written as 
\begin{equation}
  \label{forward-model-full-dataset}
\mb{d} = \diag(\mb{p}) \mb{p}^\ast \quad \text{where} \quad  \mb{p} = \mb{F} \mb{Z} \mb{a}.
\end{equation}

{ We note that the forward model in \eqref{forward-model-full-dataset} is an 
approximation to the full elastic wave equation for propagation of acoustic waves in the 
tissue. The accuracy and effectiveness of this approximation in the context of MR-ARFI 
has already been studied in detail \cite{marsac2012mr}. The main benefit
of this approximation is that the forward map can be evaluated  
efficiently which enables the use of a sampling algorithm for extraction of information 
in the Bayesian formulation.}  
\subsubsection{Relaxation to a continuous field.}
\begin{figure}[htp]
  \centering
  \begin{subfigure}{0.2\textwidth}
 \includegraphics[width =1\textwidth, clip = true, trim = 0cm 0cm 0cm
  0cm, bb = 5 0 180 250]{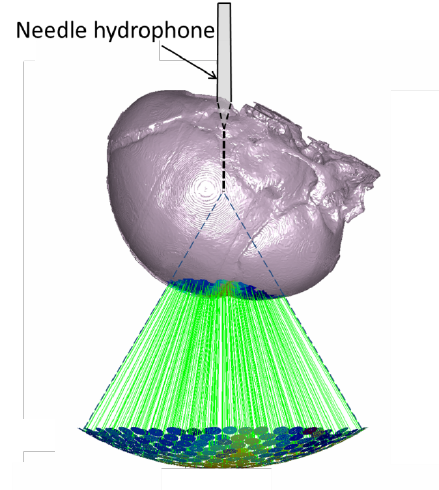} \caption{}
  \end{subfigure} 
  \begin{subfigure}{0.6 \textwidth}
  \includegraphics[width =1\textwidth, clip = true, trim = 0cm 3cm 0cm
  0cm]{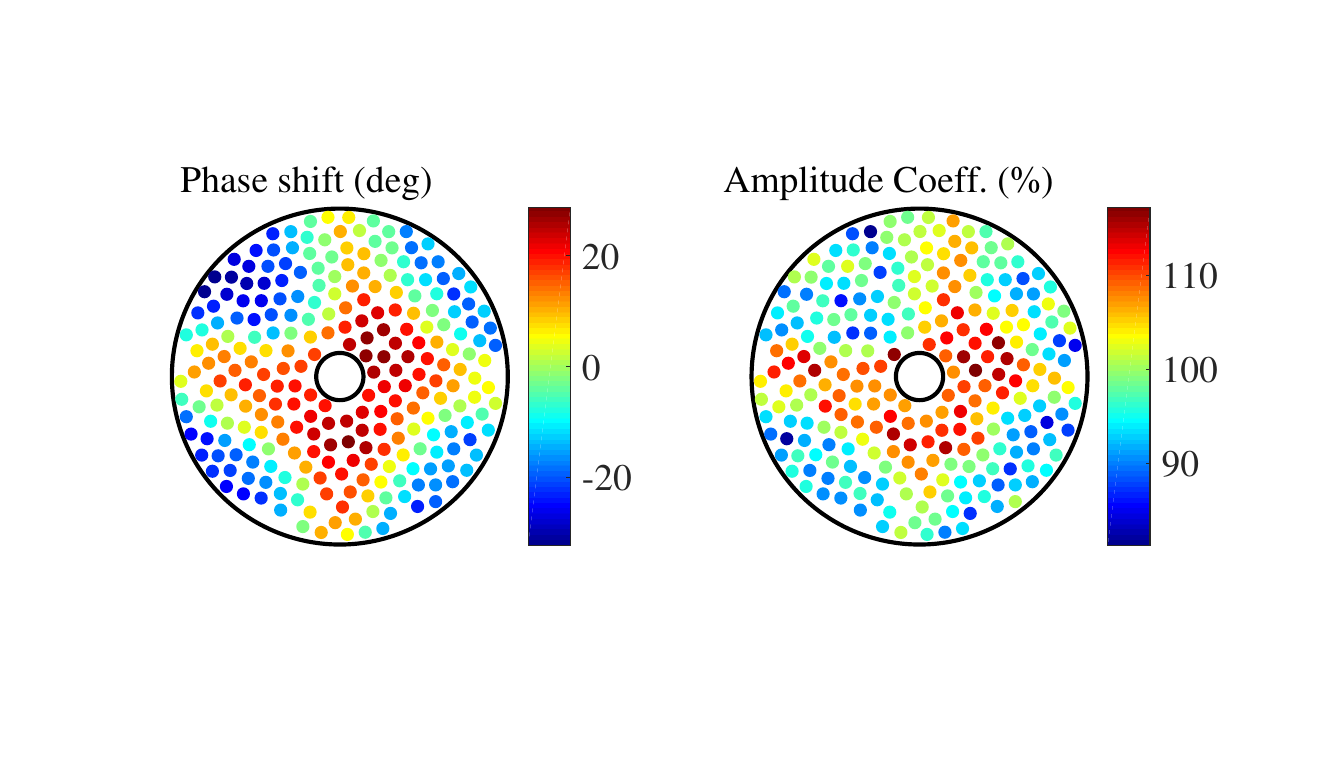}  \caption{}
  \end{subfigure}   
  \caption{(a) The setup for hydrophone
    measurements. A hydrophone is placed inside the cavity of a skull sample and is used to capture pressure signals generated by the activation of individual transducer elements. The change in phase and amplitude are calculated by comparing these signals to measurements where the skull is removed.  (b) Projection of { empirical} phase shift and amplitude  measurements 
({ corresponding to parameter $\mb{a}$ in \eqref{pressure-model})} on a newborn skull for each of the piezoelectric emitters { obtained in the article \cite{elodie}}. The aberration changes smoothly between the
    elements, which motivates the assumption that the aberrations can
    be modelled as
    a continuous function. Values of amplitude higher than 100\% are due to our normalization  so that the average of the amplitude coefficient over 
the emitters is 100\%.}
  \label{fig:skull-data}
\end{figure}

Figure \ref{fig:skull-data} 
shows examples of phase shift and attenuation
obtained with hydrophone measurements for a newborn skull using the Philips
Sonalleve V1 system 
\cite{elodie}. A schematic of the setup is depicted in Figure
\ref{fig:skull-data}(a) and
 each image (Figures \ref{fig:skull-data}(b-d)) shows the measured phase shift and
attenuation per transducer element mapped on a 2D projection of the
transducer for a different orientation of the
skull.
The aberrations appear to change
smoothly between the elements due to the presence of a soft spot in
the skull sample. This suggests that a continuous function is a good
model for the aberrations.

The acoustic
elements on the emitter are arranged on a segment of a sphere (see
Figure \ref{fig:hifu-schematic}). Project the location of the
elements on the $xy$-plane and assume that the pair $\{\bar{\mb{x}}, \bar{\mb{y}} \}$ are vectors
of the normalized $x$ and $y$ coordinates of the
elements so that all of the points fit within the unit disk in $\reals^2$. Furthermore, let $ [-1, 1]^2$ denote the
centred unit square in 2D and
consider a function $a: [-1,1]^2 \to \complex$ that
is continuous. Next, define the continuous linear
operator 
$$
S : C([-1,1]^2) \to \complex^N \quad \left(S(a;\bar{\mb{x}},
  \bar{\mb{y}})\right)_j := a(\bar{x}_{j}, \bar{y}_{j}), 
$$
for $ a \in
C([-1,1]^2)$ and $j = 1,\cdots, N$.
Combining this with \eqref{forward-model-full-dataset} defines
the forward model for a continuous aberration function:
\begin{equation} \label{forward-model-continuous}
  \mb{d} = \mcl{G}(a), \qquad \mcl{G}: C([-1,1]^2) \to \reals^{JM}, \qquad
  \mcl{G}(a) := \diag( \mb{F}\mb{Z} S(a) ) (\mb{F}\mb{Z} S(a))^\ast,
\end{equation}
where the dependence of $S$ on the coordinate
vectors is suppressed because the entries are fixed parameters that
only depend on  the geometry of the device.

\subsection{Inverse problem}
We recall 
Bayes' rule which can be written (informally) as
\begin{equation}\label{Bayes-rule}
  \pi_{\text{post}}(a | \mb{d}_{\text{obs}}) \propto \pi_{\text{likelihood}}(\mb{d}_{\text{obs}} | a) \pi_{\text{prior}}(a),
\end{equation}
where $\pi_{\text{prior}}$ is the {\it prior} probability distribution, reflecting
prior knowledge about the parameter $a$, $\pi_{\text{likelihood}}$
is the {\it likelihood} distribution, indicating the probability of an observed dataset
assuming that the parameter $a$ was known, and $\pi_{\text{post}}$ is the
{\it posterior} distribution which is the updated distribution on $a$ given
both the data and the prior. In order to avoid the delicacy of
setting up the Bayesian inverse problem on the function space we
only consider the discretized version of the problem where the
probability distributions have well-defined Lebesgue densities and refer the
reader to \cite{stuart-acta-numerica} for a detailed discussion on
this subject.

\subsubsection{Likelihood distribution $\pi_{\text{likelihood}}$}
The forward
model in \eqref{forward-model-continuous} does not incorporate
measurement errors and therefore is not a complete model
of the observed data. Assume that the error between the 
prediction of the forward model and the observed data, denoted by
$\mb{d}_{\text{obs}}$, can be modelled
as an independent Gaussian random variable. Then
\begin{equation}
\label{forward-model-statistical}
  \mb{d}_{\text{obs}} = \mcl{G}(a) + \pmb{\eta} \qquad\text{where} \qquad \pmb{\eta} \sim
  \mcl{N}(0, \pmb{\Sigma}),
\end{equation}
where $\pmb{\eta}$ is the measurement error and $\pmb{\Sigma}$
is a positive definite matrix representing the covariance of the error. The random variable $\pmb{\eta}$ has a density 
with respect to the Lebesgue measure in $\reals^{JM}$
\begin{equation}
\pi_{\pmb{\eta}}(\mb{x}) = (2 \pi)^{-\frac{JM}{2}} |\Sigma|^{-\frac{1}{2}}
\exp \left( - \frac{1}{2} \mb{x}^\ast  \pmb{\Sigma}^{-1} \mb{x}
\right).
\end{equation}
The conditional
distribution of the
observed data for a fixed $a$, which is simply the likelihood, is
given as
\begin{equation}
  \label{likelihood}
  \pi_{\text{likelihood}} (\mb{d}_{\text{obs}} | a) = \pi_{\pmb{\eta}}(
  \mb{d}_{\text{obs}}  - \mcl{G}(a) ) = \frac{1}{\beta} \exp( -\Phi(a
  ; \mb{d}_{\text{obs}})),
\end{equation}
where 
\begin{equation} \label{likelihood-potential}
  {\Phi}(a ; \mb{d}_{\text{obs}}) := \frac{1}{2} \left(
\mcl{G}(a) - \mb{d}_{\text{obs}}\right)^\ast \pmb{\Sigma}^{-1} \left(\mcl{G}(a) -
\mb{d}_{\text{obs}} \right),
\end{equation}
is referred to as the {\it likelihood potential} and $\beta := (2\pi)^{-JM/2}
|\pmb{\Sigma}|^{-1/2}$ is the normalizing constant so that
$\pi_{\text{likelihood}}$ is a proper probability distribution.

\subsubsection{Hierarchical prior distribution $\pi_{\text{prior}}$}\label{sec:hierarchical-prior}

\begin{figure}[htp]
  \centering
  \begin{subfigure}{0.44 \textwidth}
  \includegraphics[width = 1 \textwidth, clip = true, trim = 1cm 1.2cm 1cm
  1cm, bb = 30 30 1000 770]{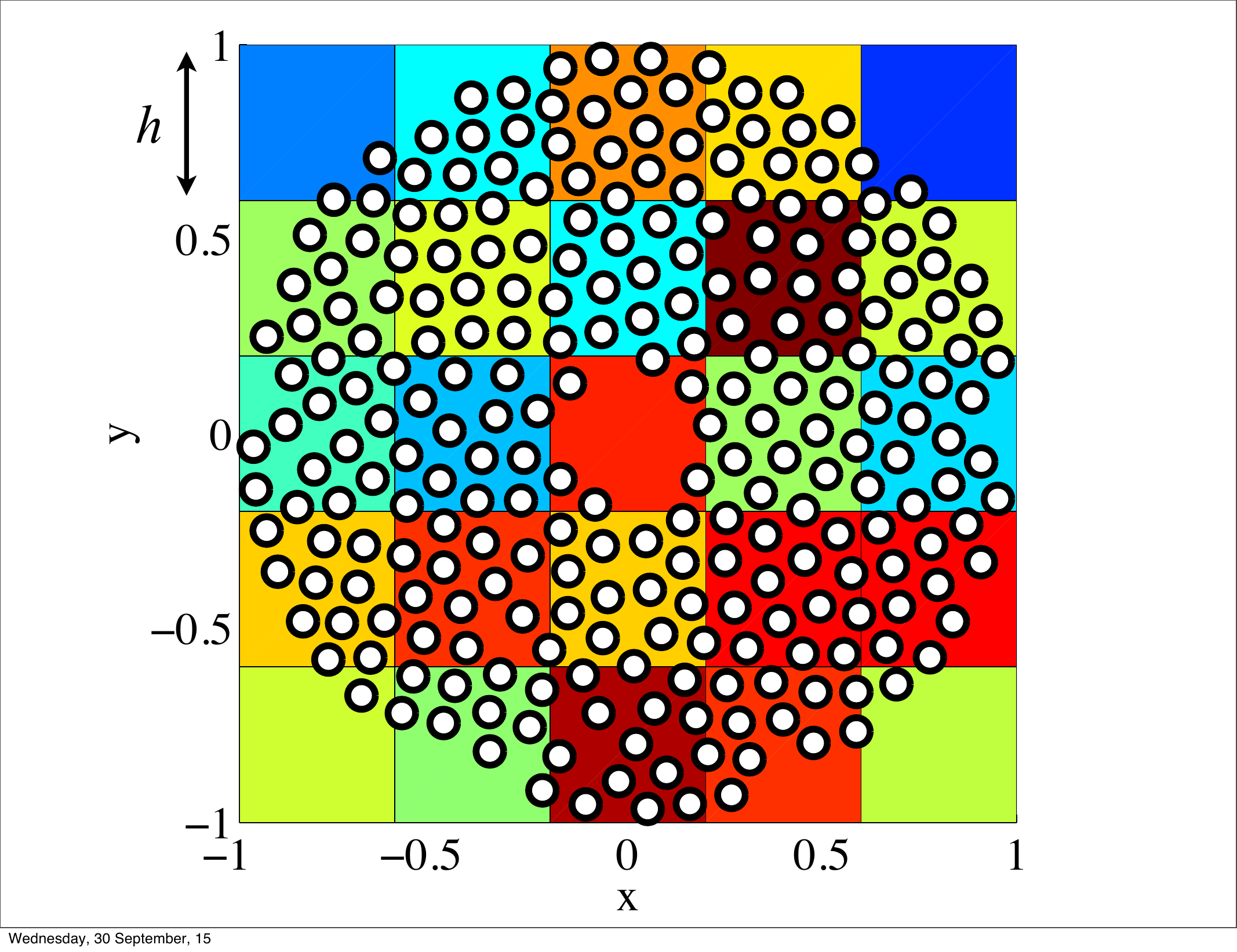}  \caption{}
  \end{subfigure}   
  \begin{subfigure}{0.55\textwidth}
\raisebox{0.5cm}{  \includegraphics[width = .98 \textwidth, clip = true, trim = 4cm 1.5cm 12cm
  0.5cm]{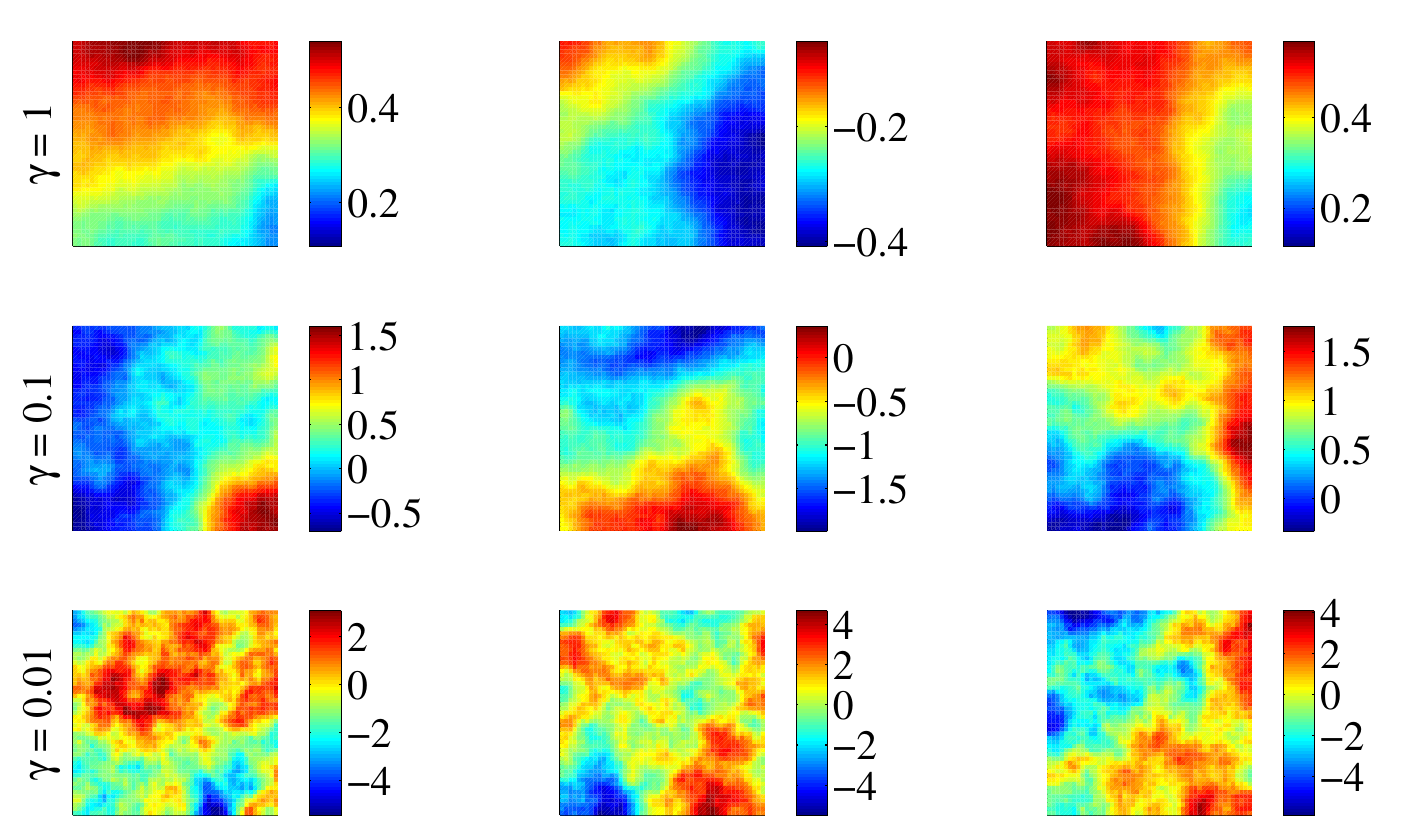}} \caption{}
  \end{subfigure} 
  \caption{(a) Schematic of the underlying field $a$ discretized on a
    uniform grid of size $h$. The value of the field is interpolated
    up to each element of the transducer using a simple interpolation matrix.
(b) Examples of draws from the Gaussian distribution $\mcl{N}(0, \mb{P}^{-1})$ for various values of
    $\gamma$ on a $50 \times 50$ grid. Smaller values of $\gamma$ result
  in samples with smaller features and larger amplitudes.}
\label{fig:prior-samples}
\end{figure}

Recall that the function $a$ was defined on the unit square $[-1,1]^2$. Now
suppose that the box is discretized using a uniform grid of size $h$ and
define the matrix $(\mb{A})_{jk} := a(-1 + (j-1)h, -1 + (k-1)h)$ for $j,k = 1,2, \cdots,
\sqrt{G}$ and $\sqrt{G} = 2/h + 1$, which is the usual finite difference discretization of
$a$ (see Figure \ref{fig:prior-samples}(a)). Now,
take $\mb{a}_h \in \complex^G$ to be the vector that is obtained by
concatenating the entries of $\mb{A}$ column by column and define
the matrices
\begin{equation}
 \tilde{\mb{L}} := \frac{1}{h^2}
\begin{bmatrix}
  2& -2 & 0 & \cdots &\cdots &0  \\
  -1 & 2 & -1 & \cdots & \cdots & 0 \\
  \vdots &  \vdots & & & & \vdots \\
  0 & \cdots  & \cdots & -1 & 2 & -1 \\
  0 & \cdots  & \cdots & & -2 & 2 \\
\end{bmatrix} \in \reals^{\sqrt{G} \times \sqrt{G}} \quad \text{and}
\quad \mb{L}:= \tilde{\mb{L}} \otimes \tilde{\mb{L}} \in
\reals^{G\times G}.
\end{equation}
The hierarchical prior distribution is constructed by first
introducing the random variables \begin{equation}\label{smoothness-prior}
\mb{u} \sim \mcl{N}(0, \mb{P}^{-1}), \quad \mb{v}
\sim \mcl{N}(0, \mb{P}^{-1}) ,\quad \alpha_1 \sim
  \mcl{N}(0, \sigma_1^2) ,\quad \alpha_2 \sim
  \mcl{N}(0, \sigma_2^2)  
\end{equation}
where $\mb{P} := h(\mb{I}_{G\times G} -
  \gamma\mb{L})^2$ and $\sigma_1$ and $\sigma_2$ are fixed. Then the
prior on the aberration is represented via the random variable
\begin{equation}\label{smoothness-prior-assembled}
\mb{a}_h \sim \pi_{\text{prior}}, \quad  
 \mb{a}_h =   \mb{a}_h(\pmb{\theta}) := \diag(\bar{\mb{u}} + 
  \alpha_1^2 \mb{u} )\exp( i ( 
  \alpha_2^2\mb{v}))
,\quad \pmb{\theta} :=
\begin{bmatrix}
  \alpha_1 &
\alpha_2 & 
\mb{u}^\ast &
\mb{v}^\ast
\end{bmatrix}.
\end{equation}
Here, $\bar{\mb{u}}$ is the prior mean of the amplification coefficient which is introduced
separately since it is fixed,
 the exponential function is applied element
by element and $\pmb{\theta}$ is introduced to simplify notation in
the next section. 

The $\mb{P}^{-1}$ covariance operator in \eqref{smoothness-prior} is a
finite difference discretization of the biharmonic operator $(I -
\gamma\Delta)^{-2}$ with homogeneous Neumann boundary conditions. Here, $\Delta$ is the
Laplacian in 2D. The finite difference matrix is scaled by a factor $h$ so that draws from
the prior have the proper white noise scaling in the continuum limit
as $h \to
0$. 
 The parameter $\gamma$ controls the size of the
features in the samples. Figure
\ref{fig:prior-samples}(b) shows a few samples of $\mcl{N}(0,\mb{P}^{-1})$
for different values of $\gamma$ discretized on a $50 \times 50$
grid (that is, $G = 50^2$). 

We refer to $\alpha_1$ and $\alpha_2$ as
hyperparameters. They control the variance of the
samples, indicating prior knowledge of the range of variations of
the phase shift or attenuation. Introducing the hyperparameters as
multipliers is crucial to making sure that the sampling algorithm in the next
section is well defined in the continuum limit \cite{agapiou}.

 It is known that in the continuum limit, draws from the Gaussian distributions in
\eqref{smoothness-prior} are almost surely Lipschitz continuous
\cite[Lemma~6.25]{stuart-acta-numerica}. Therefore, \eqref{smoothness-prior-assembled} serves as a good model for the
aberrators of Figure \ref{fig:skull-data}. 
Since the samples are continuous, a
straight forward
linear interpolation can be used to obtain point values of the samples at the
location of the elements. Therefore, the $S$ operator of
\eqref{forward-model-continuous} is easily approximated
with an interpolation matrix $\mb{S}$. 

\subsection{Sampling the posterior distribution $\pi_{\text{post}}$}
The posterior distribution $\pi_{\text{post}}$
is, in essence, the solution to the inverse problem. However,
representing this 
distribution is infeasible in practice since it may not have a closed
form. Therefore, one
often tries to
extract different statistics of this distribution or simply
obtain independent samples
from it.

Equations \eqref{likelihood} and \eqref{smoothness-prior-assembled} identify 
$\pi_{\text{post}}$ via \eqref{Bayes-rule} up to a normalizing constant. Computing this
constant is often infeasible. However,
Markov
Chain Monte Carlo (MCMC) methods can still be used to generate samples from
$\pi_{\text{post}}$
without knowledge of this normalizing constant \cite{casella}.
In this article, the preconditioned Crank-Nicolson (pCN)
and
Metropolis-Adjusted Langevin (MALA) algorithms of
\cite{stuart-mcmc} are used in a Metropolis within Gibbs (MwG) sampler
to generate samples from $\pi_{\text{post}}$. Here, the MALA algorithm is used to
sample from $\mb{u}$ and $\alpha_1$ and pCN is used to sample from
$\mb{v}$ and $\alpha_2$.
The reason for this choice is the fact that the forward map \eqref{forward-model-continuous} is not
differentiable with respect to the phase and so the pCN algorithm is
utilized to sample in this direction.
The resulting algorithm can be summarized as follows:

\paragraph{\textbf{Metropolis within Gibbs (MwG) sampler}}
\begin{enumerate}
\item Set $k=0$ and choose $\pmb{\theta}^{(0)}$ by picking $\mb{u}^{(0)}$, $\mb{v}^{(0)}$,
  $\alpha_1^{(0)}, \alpha_2^{(0)}$ randomly from the prior
  distribution and choose $\delta_1 \in [0,\infty)$ and
  $\delta_2 \in [0,1)$.
\item (MALA) Update $\mb{u}$ and $\alpha_1$:
  \begin{enumerate}[2.1.]
  \item   Propose ${\pmb{\theta}}^{(k+1/4)}$ using
$$
\begin{aligned}
& {\mb{v}}^{(k+1/4)} = \mb{v}^{(k)}, \quad\mb{w}_1 \sim \mcl{N}(0,
\mb{P}^{-1}), \\ 
& {\mb{u}}^{(k+1/4)} = \frac{2 - \delta_1}{2 + \delta_1} \mb{u}^{(k)} -
  \frac{2 \delta_1} {2 + \delta_1}\mb{P}^{-1} \nabla_{\mb{u}} \Phi \left(\mb{a}_h^{(k)};
  \mb{d}_{\text{obs}} \right) + \frac{\sqrt{8\delta_1}}{2 +\delta_1}
\mb{w_1} , \\ 
&{\alpha}_2^{(k+1/4)} = \alpha_2^{(k)}, \quad \xi_1 \sim
\mcl{N}(0,\sigma_1^2), \\
& {\mb{\alpha}}_1^{(k+1/4)} = \frac{2 - \delta_1}{2 + \delta_1} \alpha_1^{(k)} -
  \frac{2 \delta_1 c^2_1} {2 + \delta_1} \nabla_{\alpha_1} \Phi \left(\mb{a}_h^{(k)};
  \mb{d}_{\text{obs}} \right) + \frac{\sqrt{8\delta_1}}{2 +\delta_1}
\xi_1.
\end{aligned}
$$

\item Set ${\pmb{\theta}}^{(k+1/2)} = {\pmb{\theta}}^{(k+1/4)}$ with probability
$\kappa (\pmb{\theta}^{(k)} ;{\pmb{\theta}}^{(k+1/4)})$.
\item Otherwise set $\pmb{\theta}^{(k+1/2)}  = \pmb{\theta}^{(k)}$.
  \end{enumerate}
\item (pCN) Update $\mb{v}$ and $\alpha_2$:
  \begin{enumerate}[3.1.]
  \item   Propose ${\pmb{\theta}}^{(k+3/4)}$ using
$$
\begin{aligned}
&{\mb{u}}^{(k+3/4)} = \mb{u}^{(k+1/2)}, \quad\mb{w}_2 \sim
\mcl{N}(0,\mb{P}^{-1}), \\
& {\mb{v}}^{(k+3/4)} = \sqrt{1 - \delta_2^2} \mb{v}^{(k+1/2)} +
\mb{w}_2,\\
&{\mb{\alpha}}_1^{(k+3/4)} = \alpha_1^{(k+1/2)},\quad \xi_2 \sim
\mcl{N}(0,\sigma_2^2), \\
&{\alpha}_2^{(k+3/4)} = \sqrt{1-\delta_2^2} \alpha_2^{(k+3/4)} +
\xi_2.
\end{aligned}
$$

\item Set ${\pmb{\theta}}^{(k+1)} = {\pmb{\theta}}^{(k+3/4)}$ with probability
$\tau (\pmb{\theta}^{(k+1/2)} ;{\pmb{\theta}}^{(k+3/4)})$.
\item Otherwise set $\pmb{\theta}^{(k+1)}  = \pmb{\theta}^{(k+1/2)}$.
  \end{enumerate}
\item Set $k \to k+1$ and return to step 2.
\end{enumerate}
The acceptance probability in step 2.2 is defined as
\begin{equation}
  \kappa( \pmb{\theta}; \tilde{\pmb{\theta}}) := \min\left\{1,  \exp \left(
  \rho(\pmb{\theta}; \tilde{\pmb{\theta}}) -
  \rho(\tilde{\pmb{\theta}}; \pmb{\theta}) \right) \right\}
\end{equation}
where 
\begin{equation}
\begin{aligned}
\rho(\pmb{\theta}; \tilde{\pmb{\theta}}) :=  \Phi( \pmb{\theta}
; \mb{d}_{\text{obs}}) &+ \frac{1}{2} 
\begin{bmatrix}
  \tilde{\alpha}_1 - \alpha_1\\
 h^2 (\tilde{\mb{u}} - \mb{u} )
\end{bmatrix}^\ast 
\begin{bmatrix}
  \nabla_{{\alpha_1}} \Phi(\pmb{\theta}; \mb{d}_{\text{obs}})\\ 
  \nabla_{{\mb{u}}} \Phi(\pmb{\theta}; \mb{d}_{\text{obs}})
\end{bmatrix} \\
& + \frac{\delta}{4} \begin{bmatrix}
  \tilde{\alpha}_1 + \alpha_1\\
 h^2( \tilde{\mb{u}} + \mb{u} )
\end{bmatrix}^\ast 
\begin{bmatrix}
  \nabla_{{\alpha_1}} \Phi(\pmb{\theta}; \mb{d}_{\text{obs}})\\ 
  \nabla_{{\mb{u}}} \Phi(\pmb{\theta}; \mb{d}_{\text{obs}})
\end{bmatrix} \\
&  + \frac{\delta}{4} \begin{bmatrix}
  \nabla_{{\alpha_1}} \Phi(\pmb{\theta}; \mb{d}_{\text{obs}})\\ 
  \nabla_{{\mb{u}}} \Phi(\pmb{\theta}; \mb{d}_{\text{obs}})
\end{bmatrix}^\ast 
\begin{bmatrix}

\sigma_1^2 & 0 \\
0 & \mb{P}^{-1} 
\end{bmatrix} 
\begin{bmatrix}
  \nabla_{{\alpha_1}} \Phi(\pmb{\theta}; \mb{d}_{\text{obs}})\\ 
  \nabla_{{\mb{u}}} \Phi(\pmb{\theta}; \mb{d}_{\text{obs}})
\end{bmatrix}.
\end{aligned}
 \end{equation}
The acceptance probability in step 3.2 is
\begin{equation}
  \tau({\pmb{\theta}}; \tilde{\pmb{\theta}}) := \min \left\{ 1, \exp\left( 
\Phi(\pmb{\theta}; \mb{d}_{\text{obs}}) -\Phi(\tilde{\pmb{\theta}};
\mb{d}_{\text{obs}}) \right) \right\}.
\end{equation}
Derivation of the acceptance probabilities is outside the scope of
this article and the reader is referred to \cite{stuart-mcmc} for
details.  

Letting $\mb{a}_h^{(k)} := \mb{a}_h(\pmb{\theta}^{(k)})$,
the above algorithm will generate a Markov chain that has the
$\pi_{\text{post}}$ as its invariant distribution
\cite{casella}.
 This means that the samples can be used to compute the expected value of functions of the
aberrations with respect to $\pi_{\text{post}}$. Suppose that the expected
value of a function $f$ is of interest, then 
\begin{equation}
\int f(\mb{a}) \pi_{\text{post}}(\mb{a} | \mb{d}_{\text{obs}}) d\mb{a}
\approx \frac{1}{k} \sum_{\ell=1}^k f \left(\mb{a}_h^{(\ell)} \right).
\end{equation}
The functions of interest for practical
applications are the
posterior mean $\mb{a}_{\text{PM}}$, covariance $\mb{Cov} (\mb{a})$
and standard deviation $\mb{std}(\mb{a})$
of the aberration for each element:
\begin{equation}
\begin{aligned}
&  \mb{a}_{\text{PM}} \approx \frac{1}{k}
  \sum_{\ell=1}^k  \mb{S}\mb{a}_h^{(\ell)}, \\
&\mb{Cov} (\mb{a}) \approx \frac{1}{k}
  \sum_{\ell=1}^k(\mb{S}\mb{a}_h^{(\ell)} - \mb{a}_{PM})
  (\mb{S}\mb{a}_h^{(\ell)} - \mb{a}_{PM})^\ast,\\
&\mb{std}(\mb{a})_i \approx
\left( \diag(\mb{Cov}(\mb{a}))_i \right)^{1/2} \qquad \text{for}
\qquad i = 1,2, \cdots, N.
\end{aligned}
\end{equation}
Recall that $\mb{S}$ is the discrete approximation of the pointwise
evaluation operator $S$
in \eqref{forward-model-continuous} as discussed in Section~\ref{sec:hierarchical-prior}.

A key detail in implementation of the
MwG algorithm is computing the derivative of the likelihood
potential in step 2.1. This gradient can be calculated by solving an
adjoint problem. 
Let $
\mb{G} := \mb{F}\mb{Z} \mb{S} \: \diag(\exp( i \alpha_2^2\mb{v})) 
$
and recall that the pressure field can be written as
\begin{equation}
\mb{p} = \mb{F}\mb{Z}\mb{S} \: \diag (\bar{\mb{u}} + 
  \alpha_1^2 \mb{u}) \exp( i (
\alpha_2^2\mb{v})).
\end{equation}
Then, straightforward calculations show that
\begin{equation}\label{model-derivative}
 \nabla_{\alpha_1} \mathcal{G}(\pmb{\theta}) = 2\alpha_1 \diag(\mb{p})
  \mb{G}^\ast \mb{u},
\qquad
 \nabla_{\mb{u}} \mathcal{G}(\pmb{\theta})  =
2 \alpha_1^2\text{Re} \left[ \diag( \mb{p}) \mb{G}^\ast \right]. 
\end{equation}
Combining this with \eqref{likelihood-potential} gives
\begin{equation}\label{likelihood-derivative}
  \begin{bmatrix}
\nabla_{\alpha_1} \Phi(\pmb{\theta} ; \mb{d}_{\text{obs}})\\
\nabla_{\mb{u}} \Phi(\pmb{\theta} ; \mb{d}_{\text{obs}})    
  \end{bmatrix}
= 
  \begin{bmatrix}
   \nabla_{\alpha_1} \mathcal{G}(\pmb{\theta})^\ast\\
   \nabla_{\mb{u}} \mathcal{G}(\pmb{\theta})^\ast
  \end{bmatrix}
  \mb{\Sigma}^{-1}( \mathcal{G}(\pmb{\theta}) - \mb{d}_{\text{obs}}).
\end{equation}
Therefore, every step of the MALA update
costs roughly twice as much as the pCN update but MALA is more
efficient in exploring the posterior.


\section{Methods}\label{sec:methods}
In this section we describe the details of two
experiments that were performed to test our framework for estimation
of the acoustic aberrations. The first test uses a synthetic dataset of the displacement map
which is generated by the same model as the forward model of Section
\ref{section:forward-problem}. 
In the second test we use a physical
dataset that was obtained using a Philips Sonalleve V1 device.
\subsection{Test with synthetic displacement map}\label{sec:synth-disp}
The first test was performed using a synthetic dataset 
generated with the target aberrator in Figure
\ref{fig:skull-data}(b). The goal here was to test
the algorithm in a more relaxed setting where there was no discrepancy
between the forward model and the model for the data.

 Generating the synthetic
dataset involves many details including the geometry of the transducer
and the location of the focal point that are besides the main point of this
article. To keep the discussion short, we only present a 
summary of the methodology for performing the synthetic experiments.
The first step in generating the artificial dataset was to identify the
free-field matrix of the transducer $\tilde{\mb{F}}$ using 
\eqref{plane-wave} and the location of the elements and the MR-ARFI
voxels; The $k$-th column of $\tilde{\mb{F}}$ is simply the free-field corresponding to
the contribution of the $k$-th ultrasound emitter { evaluated at the center of 
each voxel in the MR-ARFI images}. Afterwards, the matrix ${\mb{F}}$ is constructed using
\eqref{transducer-formulation}.
The next step was to construct
the design matrix $\mb{Z}$ which is identified by the
$\tilde{\mb{z}}_j$ vectors. This matrix contains the prescribed values of
the amplitude and phase of the acoustic waves at the transducer. 
The virtual elements of \cite{herbert-energy}
were used to group nearby piezoelectric emitters and construct the design matrix.
 Let $\mb{H}_{256}$ denote the $256
\times 256$ Hadamard matrix \cite[Eq.~8]{herbert-energy} with columns
$\mb{h}_1$
to $\mb{h}_{256}$. Then $\tilde{\mb{z}}_j = \mb{h}_1 +
\exp(i \frac{\pi}{6})\mb{h}_j $ for $j = 2,\cdots, 16$. 

\begin{figure}[htp]
  \centering
  \begin{overpic}[width =1 \textwidth, clip=true, trim = 5cm 2cm 5cm 1cm]{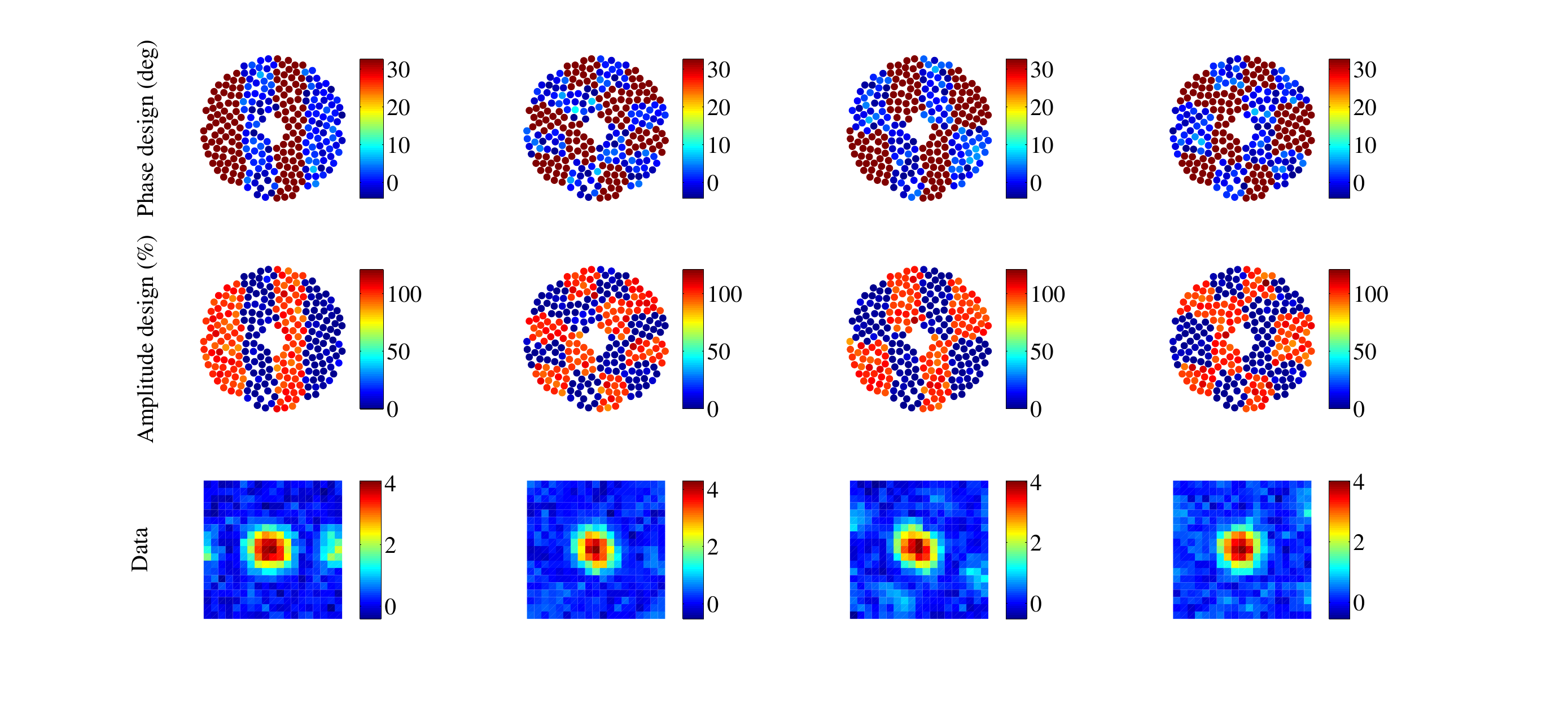}
    \put(0,16){\rotatebox{90}{\tiny Amplitude design (\%)}}
    \put(0,35){\rotatebox{90}{\tiny Phase design (deg)}}
    \put(0,1){\rotatebox{90}{\tiny Displacement map}}
\end{overpic}
  \caption{A few examples of artificial experiments, indicating the
    prescribed noisy phase
    and amplitude (the $\tilde{\mb{z}}_j^{\text{noisy}}$ vectors) along with the displacement map (the resulting MR image) of the focal
    point (subsets of $\mb{d}_{\text{obs}}^{\text{art}}$) scaled by the standard deviation of the images. The displacement maps show the focal point before refocusing. 
   }
  \label{fig:Hadamard-basis}
\end{figure}

A noisy version of the design
vectors was considered with
\begin{equation}
\tilde{\mb{z}}_{j}^{\text{noisy}} := \tilde{\mb{z}}_j +
\pmb{\epsilon}_1 \exp\left(i \frac{\pi}{4} \pmb{\epsilon}_2\right) \quad \text{where}
\quad \pmb{\epsilon}_1, \pmb{\epsilon}_2 \sim \mcl{N}(0, (0.05)^2 \cdot\mb{I}_{256 \times
  256}).
\end{equation}
This added an extra layer of noise that is not accounted for in the
formulation of the
inverse problem. The standard deviation of the noise in the phase was
taken to be $0.05$ as a reasonable estimate of the errors in imposing
the sonication patterns in practical settings.
Putting these vectors together gave a noisy design matrix
$\mb{Z}^{\text{noisy}}$ and the synthetic dataset was
generated using
\begin{equation}
\mb{d}_{\text{obs}}^{\text{art}} = \mb{F}
\mb{Z}^{\text{noisy}} \mb{a}^{\text{art}} + \pmb{\eta} \quad \text{where}
\quad \pmb{\eta} \sim  \sigma_{\text{obs}}^{\text{art}} \cdot \mcl{N}(0,
(0.2)^2 \cdot\mb{I}_{256 \times 256})
\end{equation}
and $\sigma_{\text{obs}}^{\text{art}}$ is the standard deviation of
$\mb{F}\mb{Z}^{\text{noisy}} \mb{a}^{\text{art}}$. The standard
deviation of the measurement noise was taken to be $0.2$ which implied 
a signal to noise ratio of $5$. This choice was made to replicate data
that is highly noisy. 
 Figure
\ref{fig:Hadamard-basis} shows examples of the generated noisy data
along with the prescribed phase and amplitudes of the elements.

A total of 16 sonication tests were performed in this example, that is, the design matrix
had $16$ columns.
In order to solve the inverse problem, a coarse $8 \times 8$ mesh
($h = 1/4$) was used to
discretize the aberration field $a$. It was assumed that the noise
covariance $\pmb{\Sigma} = (0.15 \cdot
\sigma_{\text{obs}}^{\text{art}})^2 \cdot \mb{I}_{JM
  \times JM}$.
 The prior mean was taken to be $\bar{\mb{u}}=1$, the prior wave number $\gamma=1/5$ and the
hyperparameter prior variances were $\sigma_1 = \sigma_2 =2$. 
The step sizes in the MCMC algorithm were $\delta_1 = 4
\times 10^{-5}$ and $\delta_2 = 5 \times 10^{-3}$ resulting in
an average acceptance
probability of $0.32$ across the two Metropolis Hastings updates.



\subsection{Test with MR-ARFI displacement map} \label{methods:physical-test}

A test was done using the displacement maps obtained from MR-ARFI data
which was acquired
using a Philips Sonalleve V1 ultrasound system 
(Philips Healthcare, Vantaa, Finland)
and an
Achieva 3T MRI machine (Philips, Best, Netherlands). A phased array transducer consisting
of 256 elements was used with a focal
length of 12 cm and  aperture of 13 cm. 
 The transducer was operating at 1.2
MHz generating ultrasound pulses of 1 ms ranging from 0
to 300 acoustic Watts. The transducer was submerged in an oil tank (as
illustrated in Figure \ref{fig:hifu-schematic}) and the acoustic field
was targeted at a 4 cm phantom which mimics muscle tissue (CIRS,
Norfolk, VA, USA). 
The displacement due to the ultrasound pulses was measured using a
dedicated MR receiver coil of 4 cm inner diameter. The displacements
were measured using MR-ARFI sequences based on a RF-spoiled
gradient-recalled
 Echo-Planar Imaging (EPI) sequence with Repetition Time (TR) of $42$
 ms, Echo Time (TE) of $30$ ms, flip
 angle of $20$ degrees, EPI factor of 9, MRI resolution of $64
 \times 64$ pixels,
 bi-polar motion encoding gradients of 1 ms in duration and amplitude
 of 30 mT/m. Images were obtained  every 0.26 seconds from a single slice that was
 orthogonal to the path of the beam with a field of view of $180
 \times 180$ mm$^2$ and voxels of size $0.7 \times 0.7 \times 4$ mm$^3$.

At first, a set of measurements was taken without an aberrator. 
This dataset is used for estimating the free-field matrix of the
transducer and so it is referred to as the calibration dataset.
To obtain this dataset the phase stepping technique of \cite{herbert-energy,
  larrat} was used with $16$ virtual elements (the first $16$ columns
of the Hadamard matrix) with ten steps in phase for each virtual
element. Thus, this dataset consists of $160$ sonication patterns.
 A cosine function was then fit to the displacement of
each voxel during the phase stepping in order to retrieve the phase and estimate the 
free-field matrix of each element. 

Afterwards, a second set of measurements was taken in the presence of the
 aberrator of Figure \ref{fig:skull-data}(a)
 consisting of $32$ sonication patterns.
This dataset is referred to as the
 reconstruction dataset.
The input phase and amplitude of
the 256 elements were modified to imitate the effect of the target aberrator instead of
using a physical aberrator in front of the beam. That is, the
aberrator was added to the prescribed excitation patterns during each
sonication test. This approach is
advantageous because the values of the aberration parameters are known
and so we can easily assess the quality of the estimates.
Programming of the HIFU system and collection of MR images  were performed using the toolboxes MatHIFU and MatMRI 
\cite{zaporzan2013matmri}.
Here, the design matrix $\mb{Z}$ was constructed from the vectors $\tilde{\mb{z}}_j =\frac{1}{2} ( \mb{h}_1 +
\mb{h}_j )$ for $j=1,\cdots,16$ and $\tilde{\mb{z}}_j =\frac{1}{2} ( \mb{h}_1 +
\exp(i \frac{\pi}{3})) \mb{h}_j )$ for $j = 17,\cdots,32$.
All tests were performed with a
field of view of $7 \times 7$ voxels, corresponding to $\sqrt{M} = 7$. Only a small portion of
the images are used from each frame since signal to noise ratio drops
rapidly for voxels that are far from the focal point. Each measurement was repeated
ten times and then
averaged in order to reduce the
noise. 

\bhedit{A consequence of the phase stepping method is that 
displacement maps are normalized and so the dataset does not include 
any information regarding the amplitude coefficients. To this end,  
the inverse problem is only solved for the phase shift and hyperparameter $\alpha_2$.}   
The noise covariance in the formulation of the inverse problem
 was taken to be $\pmb{\Sigma} =  ( 0.2 \cdot \sigma_{\text{obs}})^2 \cdot \mb{I}_{JM
  \times JM}$ where $\sigma_{\text{obs}}$ is the standard deviation
of the reconstruction dataset. Recall that here, $M = 49$ (number of voxels) and $J =
16$ (number of images). As before, an $8 \times 8$ grid was used for
discretization of the aberration field $a$. The prior mean
was taken to be $\bar{\mb{u}} =1$ and the prior wave
number $\gamma = 1/5$. {The prior variances on the hyperparameters were
$\sigma_1= 0.5$ and $\sigma_2=0.5$ and the MCMC step sizes were $\delta_1
=1.1 \times 10^{-5}$ and  $\delta_2 =1.5 \times 10^{-3}$ resulting in an average acceptance probability of
$0.54$ across the two Metropolis Hastings updates.} \bhedit{Since
inversion is performed solely for the phase shift, only the pCN update was utilized.}

\subsubsection{Calibration of the free-field matrix}\label{sec:calibration}
The free-field matrix of the transducer was computed 
using the phase stepping technique of \cite{herbert-energy, larrat} but
the 
estimated field is often not accurate enough to give a satisfactory
estimate of the aberrations. This issue is amplified when measurements
are noisy and the number of sonication tests is significantly
smaller 
than the number of elements on the transducer.
Furthermore, the
MR-ARFI data consists of measurements of displacement while the
forward model of Section~\ref{section:forward-problem} is valid for acoustic intensity. Although
acoustic intensity is expected to be proportional to displacement
\cite{maier}, the constant of
proportionality is unknown.

These
discrepancies will manifest as an apparent aberrator in front of the
beam. For example, running the algorithm on the
calibration dataset would still estimate a significant value for
the aberrations. In practice this aberrator must be estimated in a 
calibration step before computing the actual aberrator using the 
reconstruction dataset. This will also automatically estimate the constant of
proportionality between intensity and displacement.
 Here, the posterior mean of this inherent aberrator, denoted by
 $\mb{a}_{\text{calibration}}$, is computed  over
the first 32 measurements of the calibration dataset.
Once this vector is available, the calibration can be performed by
simply replacing the matrix
$\mb{S}$ by 
\begin{equation}
  \mb{S}_{\text{calibrated}} = \diag( \mb{a}_{\text{calibration}}) \mb{S}. 
\end{equation}

\subsection{Assessing the quality of refocusing}
Take the posterior mean $\mb{a}_{\text{PM}}$ to be a good estimator of the true aberration $\mb{a}$,
and let $\pmb{\phi}_{\text{PM}}$ and $\pmb{\phi}$ be their
corresponding phase shift vectors. 
Define
the vectors 
$$
\begin{aligned}
&\mb{e}_1 := \diag[ \mb{\tilde{F}}  \exp(i (
\pmb{\phi} - \pmb{\phi}_{\text{PM}}))] \: \mb{\tilde{F}}  \exp(i
(\pmb{\phi} - \pmb{\phi}_{\text{PM}}) ),\\
&\mb{e}_2 := \diag[ \mb{\tilde{F}}  \exp(i 
\pmb{\phi})] \: \mb{\tilde{F}}  \exp(i
\pmb{\phi} ),  \\ 
&\mb{e}_3 := \diag[ \mb{\tilde{F}} \mb{1}] \: \mb{\tilde{F}}\mb{1},
\end{aligned}
$$
and the
expected improvement $EI$ and expected recovery $ER$ functionals
 for the posterior mean
$$
EI(\mb{a}_{\text{PM}}) := \left( 1 - \frac{\| \mb{e}_3 \|_\infty - \|\mb{e}_1\|_\infty}{\| \mb{e}_3 \|_\infty - \|\mb{e}_2\|_\infty} \right)\times 100, \quad 
ER(\mb{a}_{\text{PM}}) := \left( \frac{\| \mb{e}_1 \|_\infty - \|
    \mb{e}_2\|_\infty} { \| \mb{e}_3\|_\infty }\right)
\times 100.
$$
$EI$ measures the
percentage of lost intensity that is recovered while $ER$ measures how the maximum intensity of the refocused beam compares to the 
the maximum intensity of a perfectly focused beam.
\bhedit{Since power is directly related to beam intensity, the expected improvement $EI$ can be 
viewed as a measure of improvement in treatment efficiency. Furthermore,
based on Pennes bio-heat law \cite{pennes1948analysis}, beam intensity is proportional to 
the temperature increase at the focal point. This temperature increase has an exponential relationship 
to the tissue damage based on thermal dose model \cite{sapareto1984thermal}.  
Then the expected recovery $ER$ can be viewed as a measure of improvement in 
required
treatment dosage after refocusing.}
These measures are used in the Section~\ref{sec:results}  to further assess the
performance of the estimated aberrations.

\section{Results}\label{sec:results}


\subsection{Test with synthetic displacement map}
\begin{figure}[htp]
  \centering
  \begin{subfigure}[b]{0.23\textwidth}
    \begin{overpic}[width = 1 \textwidth, clip=true, trim = 4.5cm 1cm 7.8cm
      0cm]{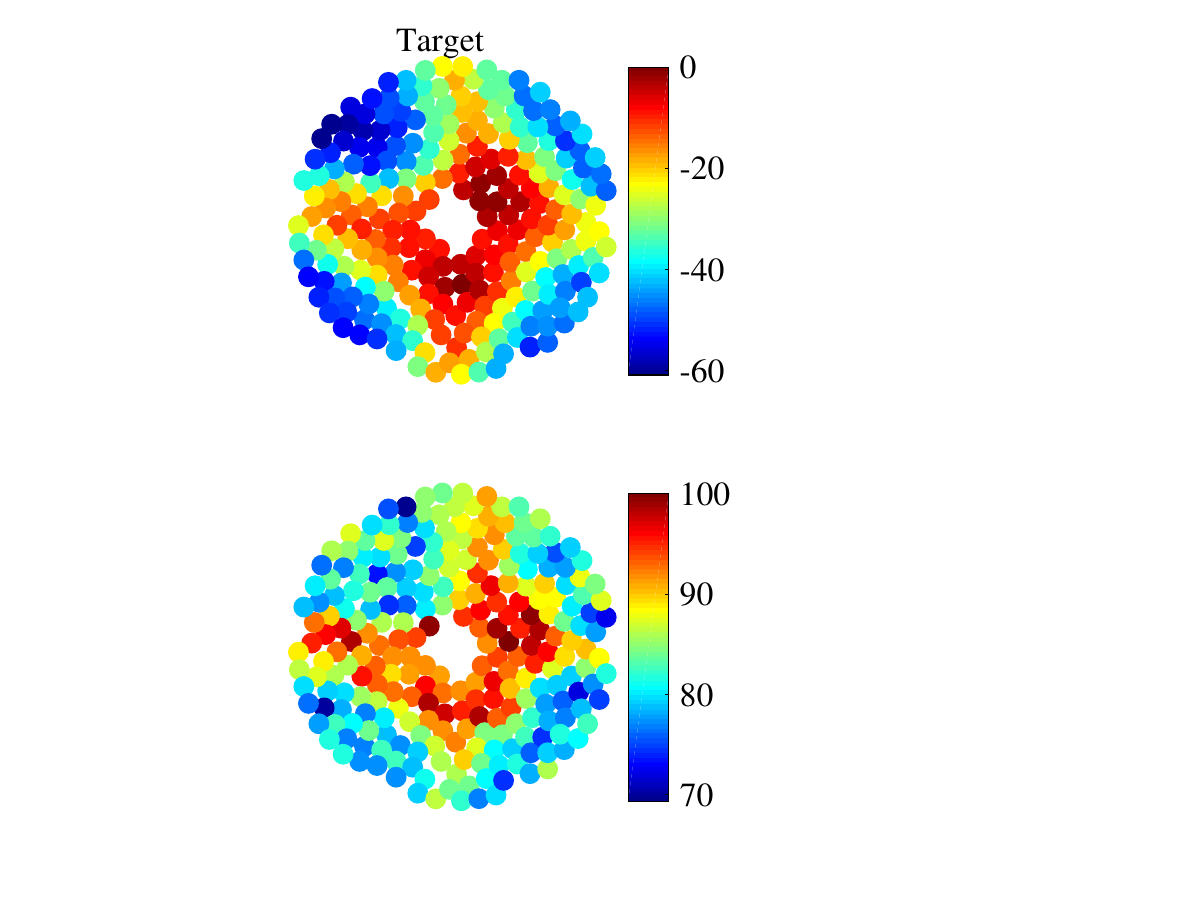}
      \put(-5,58){\rotatebox{90}{\scriptsize Phase shift (deg)}}
      \put(-5,0){\rotatebox{90}{\scriptsize Amplitude coeff. (\%)}}
    \end{overpic}

 \caption{}  
  \end{subfigure}
  \begin{subfigure}[b]{0.23\textwidth}
  \includegraphics[width =1 \textwidth, clip=true, trim =  4.5cm 1cm 7.8cm
  0cm]{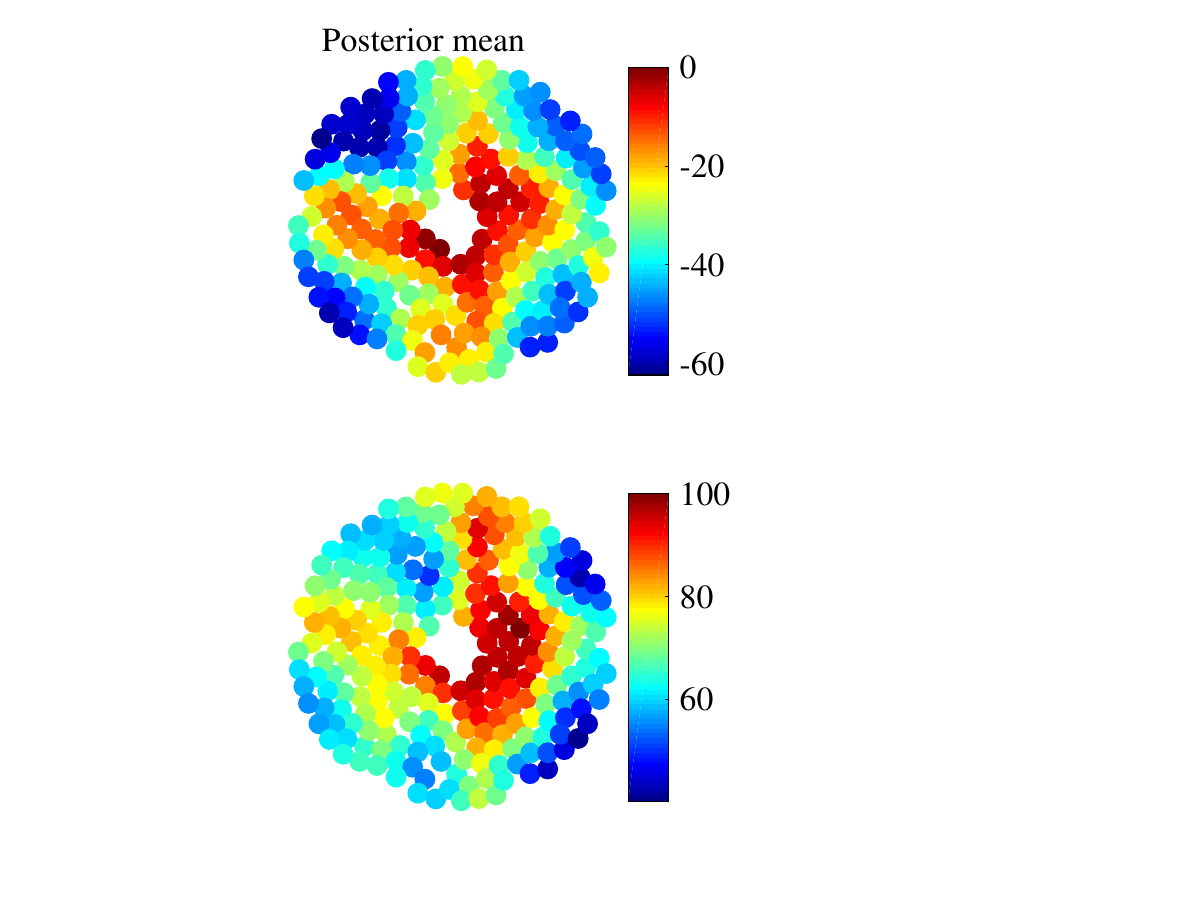}
 \caption{}  
  \end{subfigure}
  \begin{subfigure}[b]{0.23\textwidth}
  \includegraphics[width = 1 \textwidth, clip=true, trim = 4.5cm 1cm 7.8cm
  0cm]{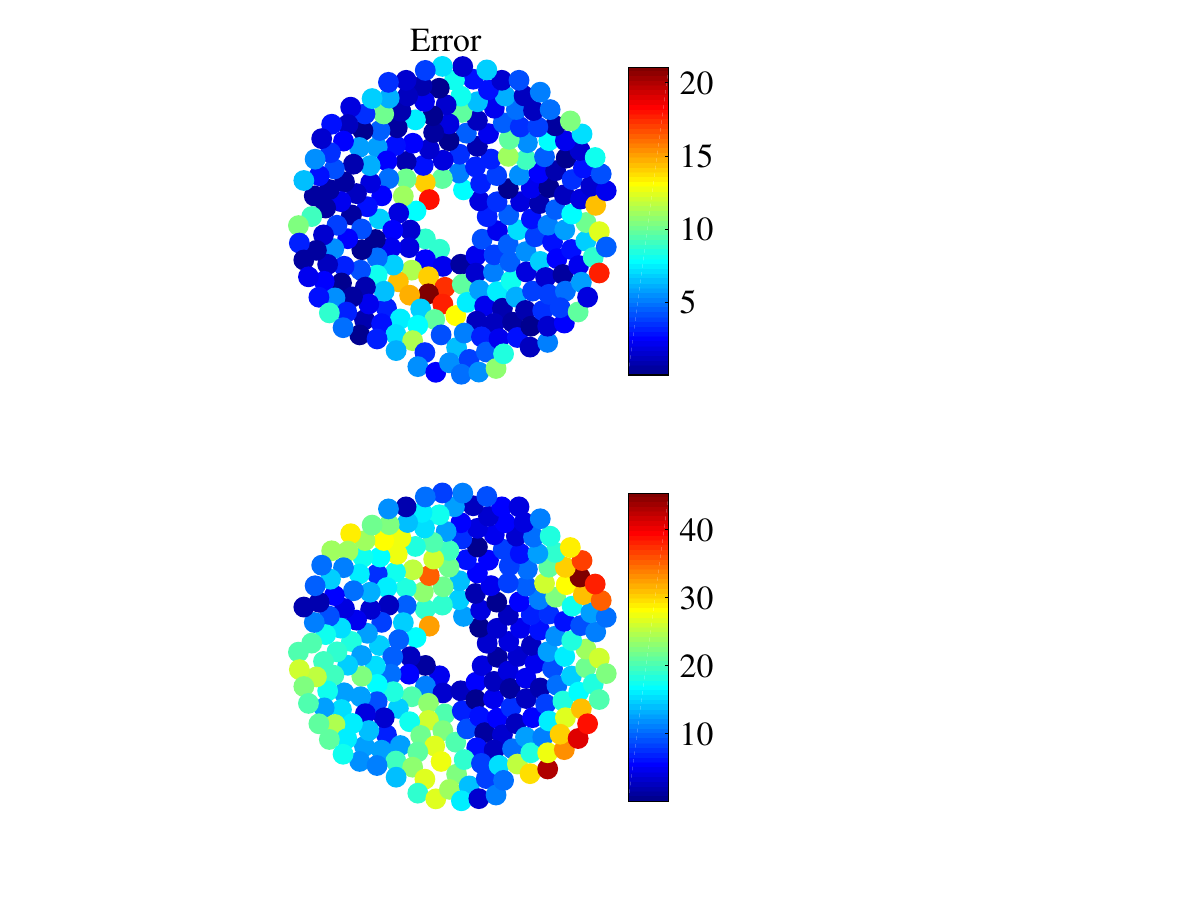}
 \caption{}  
\end{subfigure}
  \begin{subfigure}[b]{0.23\textwidth}
  \includegraphics[width = 1\textwidth, clip=true, trim =  4.5cm 1cm 7.8cm
  0cm]{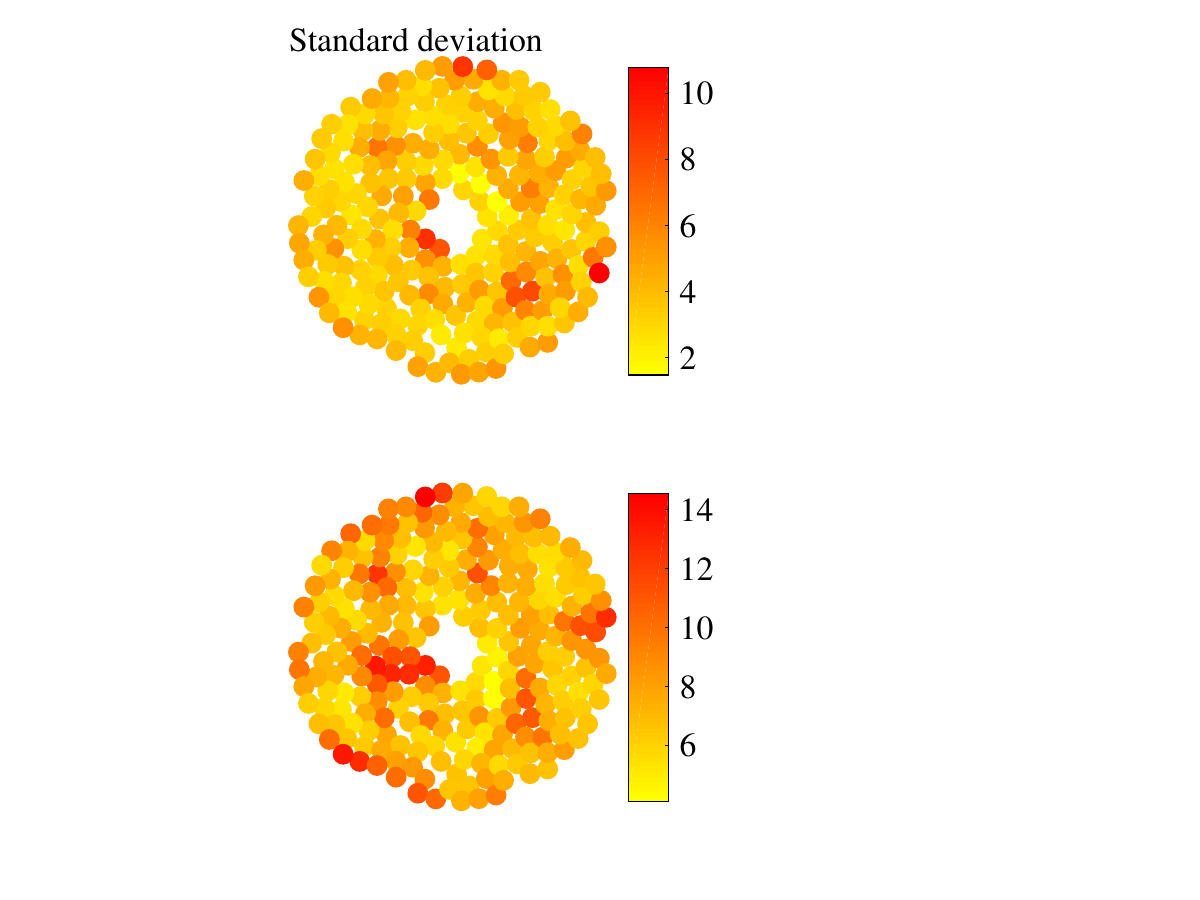}
 \caption{}  
\end{subfigure}\\
  \begin{subfigure}[b]{0.2 \textwidth}
  \includegraphics[width = 1.1 \textwidth, clip=true, trim = 0cm 0cm 0cm
  0cm]{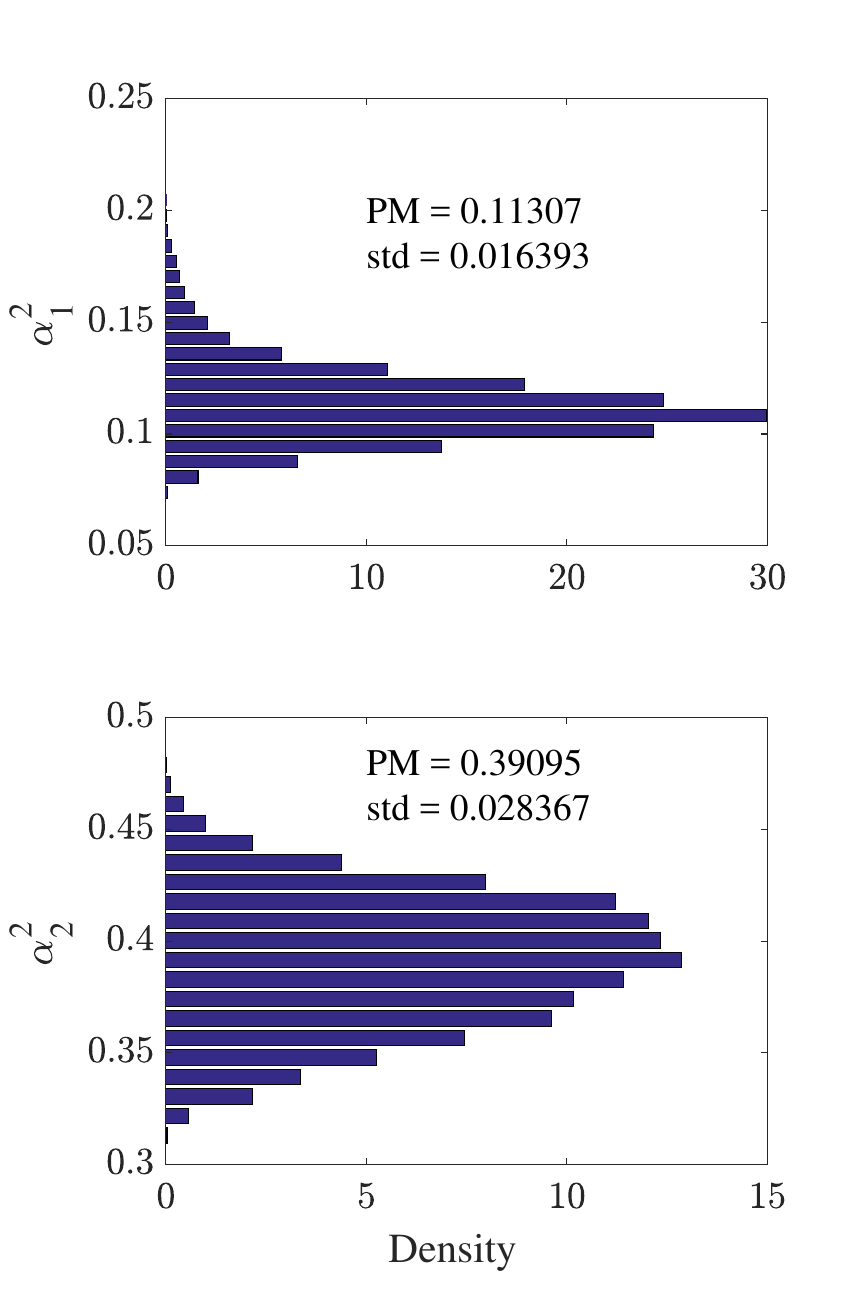}   
  \caption{}
  \end{subfigure}
  \begin{subfigure}[b]{0.75\textwidth}
    \qquad\begin{overpic}[width = 0.27 \textwidth, clip=true, trim = 3cm 0cm 0cm
  0cm]{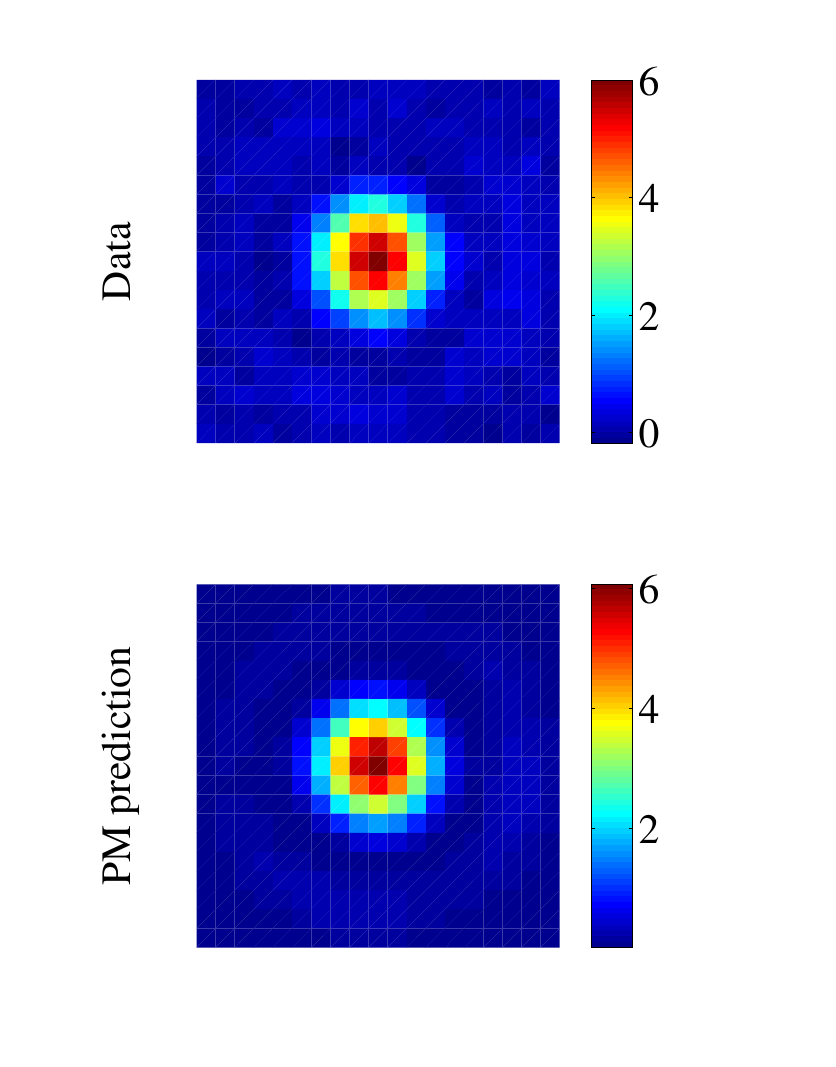}
      \put(-5,56){\rotatebox{90}{\tiny Displacement map}}
      \put(-5,12){\rotatebox{90}{\tiny PM prediction}}
    \end{overpic}
  \includegraphics[width = 0.32 \textwidth, clip=true, trim = 1cm 0cm 0cm
  0cm]{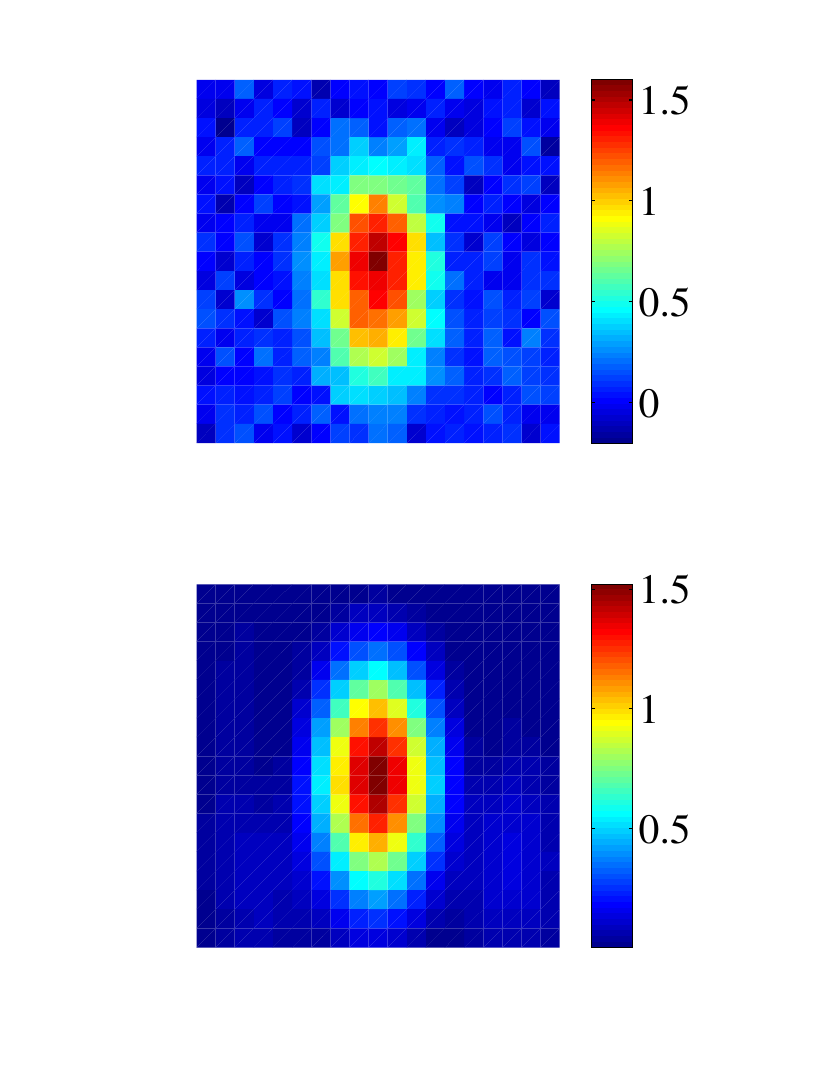}
  \includegraphics[width = 0.32 \textwidth, clip=true, trim = 1cm 0cm 0cm
  0cm]{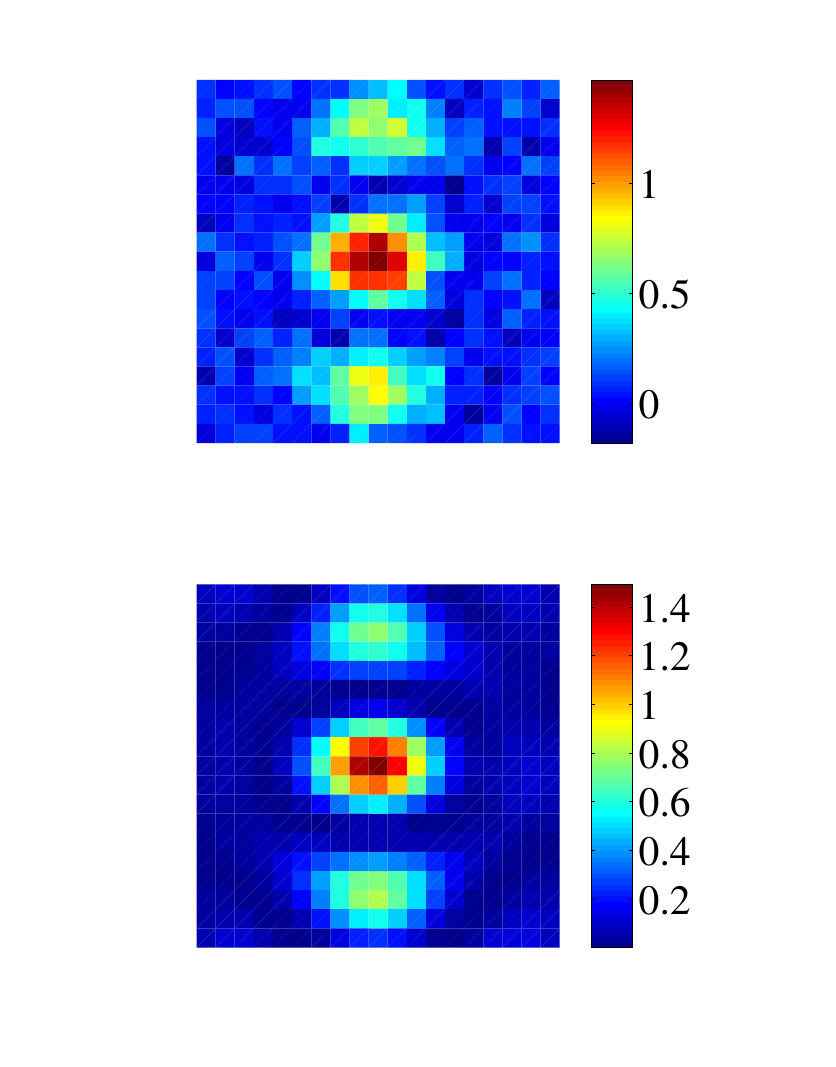}
    \caption{}
  \end{subfigure}
  \caption{A summary of Bayesian posterior statistics using the synthetic
    dataset. (a) The target aberrator used for generating the
  data. (b) Posterior mean (PM) of the aberration parameters along
  with (c) the pointwise difference between the posterior mean and the
  target aberrator.
 (d) The standard deviation (std) of the posterior samples which is indicative of the
level of uncertainty. (e) Estimated marginal posterior distributions
on the hyperparameters. (f) A few examples of the synthetic displacement 
map at the focal point compared with the prediction of 
the posterior mean of the aberrator.}
  \label{fig:artificial-CM-std-reconstruction}
\end{figure}

A summary of Bayesian posterior statistics using the synthetic dataset is
presented in Figure \ref{fig:artificial-CM-std-reconstruction}.
Posterior mean and standard deviations are computed using $5 \times
10^5$ samples with a burn-in period of $3 \times 10^3$ samples
(that is, the first $3\times 10^3$ samples were discarded since the Markov
chain had not yet converged at that point). 
The posterior mean (Figure
\ref{fig:artificial-CM-std-reconstruction}(b)) is taken to be a good estimator of the actual value
of the parameters. This is supported by Figure
\ref{fig:artificial-CM-std-reconstruction}(c) which is the pointwise
absolute difference between the posterior mean and the target
aberrator. Here the maximum error in the phase is $21$ degrees while
the average error (among the elements) is $4.5$ degrees. As for the attenuation, the maximum
error is $45$ percent and the average error is $14$ percent.
Compare these values to Figure
\ref{fig:artificial-CM-std-reconstruction}(d) which depicts the standard
deviation of the aberration parameters and can be taken as a measure
of uncertainty in the approximations. The standard deviation of the
aberrations is close the the average point wise absolute error.
Therefore, the standard deviation is a good measure of the accuracy of
$\mb{a}_{\text{PM}}$.
Furthermore, Figure \ref{fig:artificial-CM-std-reconstruction}(f)
compares the prediction of the data using $\mb{a}_{\text{PM}}$ to the actual data set
on a few frames of the MR-ARFI data. This shows a good agreement
between the prediction and the data and further certifies the choice
of $\mb{a}_{\text{PM}}$ as a pointwise estimator of the target. Finally,
 the expected improvement $EI(\mb{a}_{\text{PM}}) \approx
71 \%$ and $ER(\mb{a}_{\text{PM}}) \approx 5 \%$.  Therefore, using
$\mb{a}_{\text{PM}}$ to refocus the beam recovers $71 \%$ of the
lost intensity. However, this only improves maximum intensity
by  $5\%$ which is due to the fact that the aberrator of
Figure \ref{fig:artificial-CM-std-reconstruction}(a) is very weak
and the defocused beam is already at $93\%$ intensity.

In the case of the hyperparameters, one can integrate out the rest of
the parameters and directly estimate the marginal posterior
distribution as in Figure
\ref{fig:artificial-CM-std-reconstruction}(e). 
\bhedit{Here we demonstrate the marginals on $\alpha_1^2$ and $\alpha_2^2$ rather than 
$\alpha_1$ and $\alpha_2$
 as these  are the standard deviations of the fields $\mathbf{u}$ and $\mathbf{v}$
and are more physically relevant.}
These results indicate
a drop of two orders of magnitude in the standard deviation of the hyperparameters, compared with the prior standard deviation, indicating that the value of the hyperparameters are
computed with high confidence.

\begin{figure}[htp]
  \centering
  \begin{overpic}[width = 1 \textwidth, clip=true, trim = 5.5cm 1cm
    4.5cm .5cm]{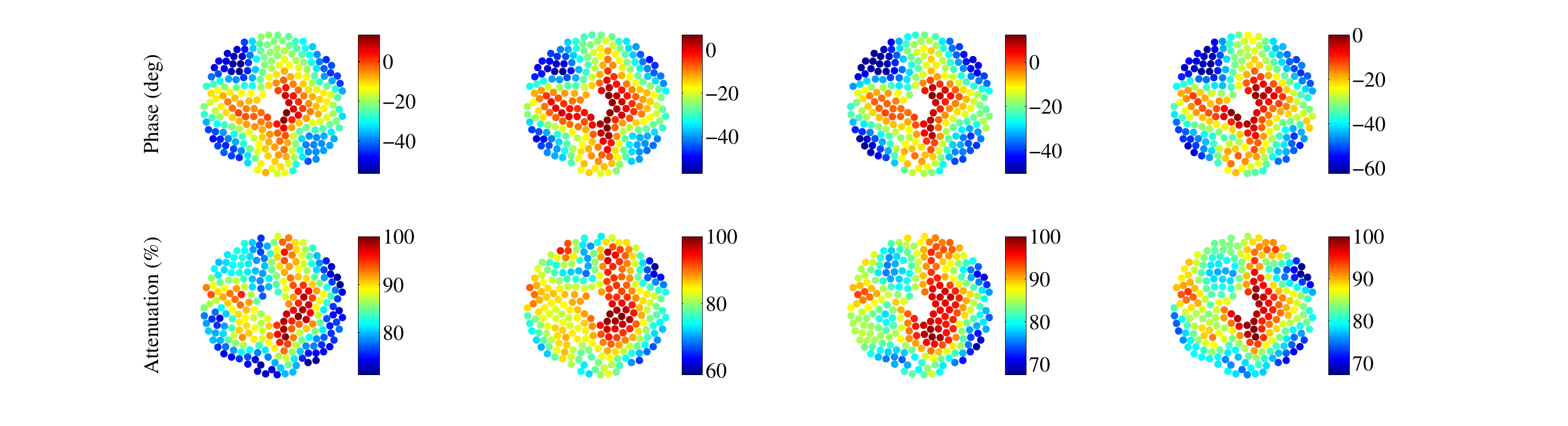}
    \put(0,0){\rotatebox{90}{\tiny Amplitude coeff. (\%)}}
    \put(0,17){\rotatebox{90}{\tiny Phase shift (deg)}}
  \end{overpic}

  \caption{A few samples from  $\pi_{\text{post}}$ on the
    aberration parameters. The samples are $10^5$ steps
    apart in the Markov chain so that they can be treated as independent.}
  \label{fig:artificial-Posterior-samples}
\end{figure}

In addition to the above statistics, one can also look at independent samples
from $\pi_{\text{post}}$ as depicted in Figure
\ref{fig:artificial-Posterior-samples}. 
These samples are generated
by choosing individual samples from the Markov chain that are far enough apart. 
The distance between the samples in Figure
\ref{fig:artificial-Posterior-samples} was
 chosen large enough so that the 
integrated autocorrelation function of the chain was below $10^{-3}$. In this
case the distance was taken to be $10^5$ steps based
on the worst integrated autocorrelation function in Figure
\ref{fig:MCMC-performance}(a). The independent samples can
be taken as examples of aberrators that are likely to have generated
the dataset and provide further insight regarding
$\pi_{\text{post}}$. It is clear that the samples have very similar features
in comparison to the mean. For example, there are no discontinuities or
multi-modal behavior. This further supports the choice of the posterior
mean $\mb{a}_{\text{PM}}$ as a point estimator for the true value of
the parameters.

\subsection{Test with MR-ARFI displacement map}
\begin{figure}[htp]
  \centering
  \begin{subfigure}[b]{0.23\textwidth}
 \begin{overpic}[width = 1 \textwidth, clip=true, trim = 4.5cm 8cm 7.8cm
      0cm]{./figs/artificial_phase_target.pdf}
      \put(-5,15){\rotatebox{90}{\scriptsize Phase shift (deg)}}
    \end{overpic}
 \caption{}  
  \end{subfigure}
  \begin{subfigure}[b]{0.23\textwidth}
  \includegraphics[width = 1 \textwidth, clip=true, trim = 4.5cm 8cm 7.8cm
  0cm]{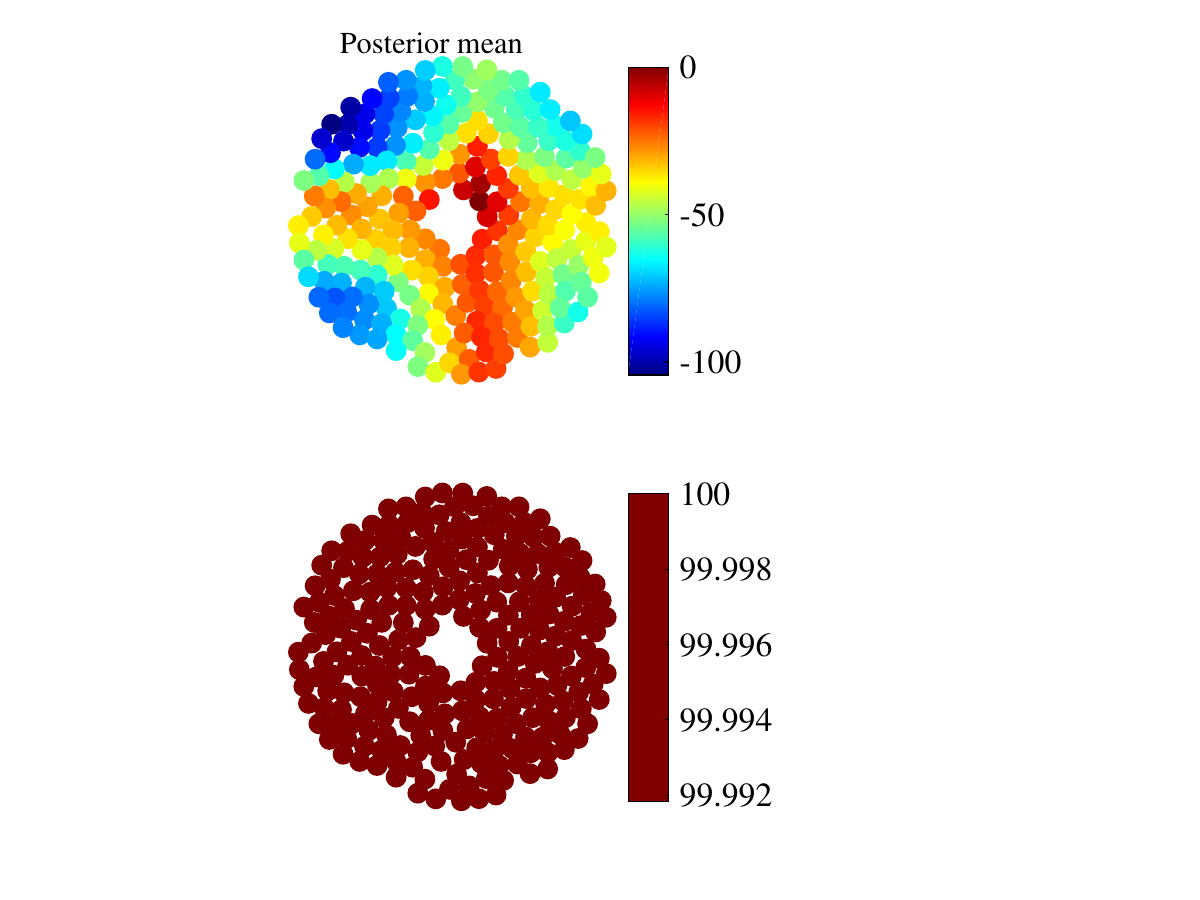}
 \caption{}  
  \end{subfigure}
  \begin{subfigure}[b]{0.23\textwidth}
{\includegraphics[width = 1 \textwidth, clip=true, trim = 4.5cm 8cm 7.8cm
  0cm]{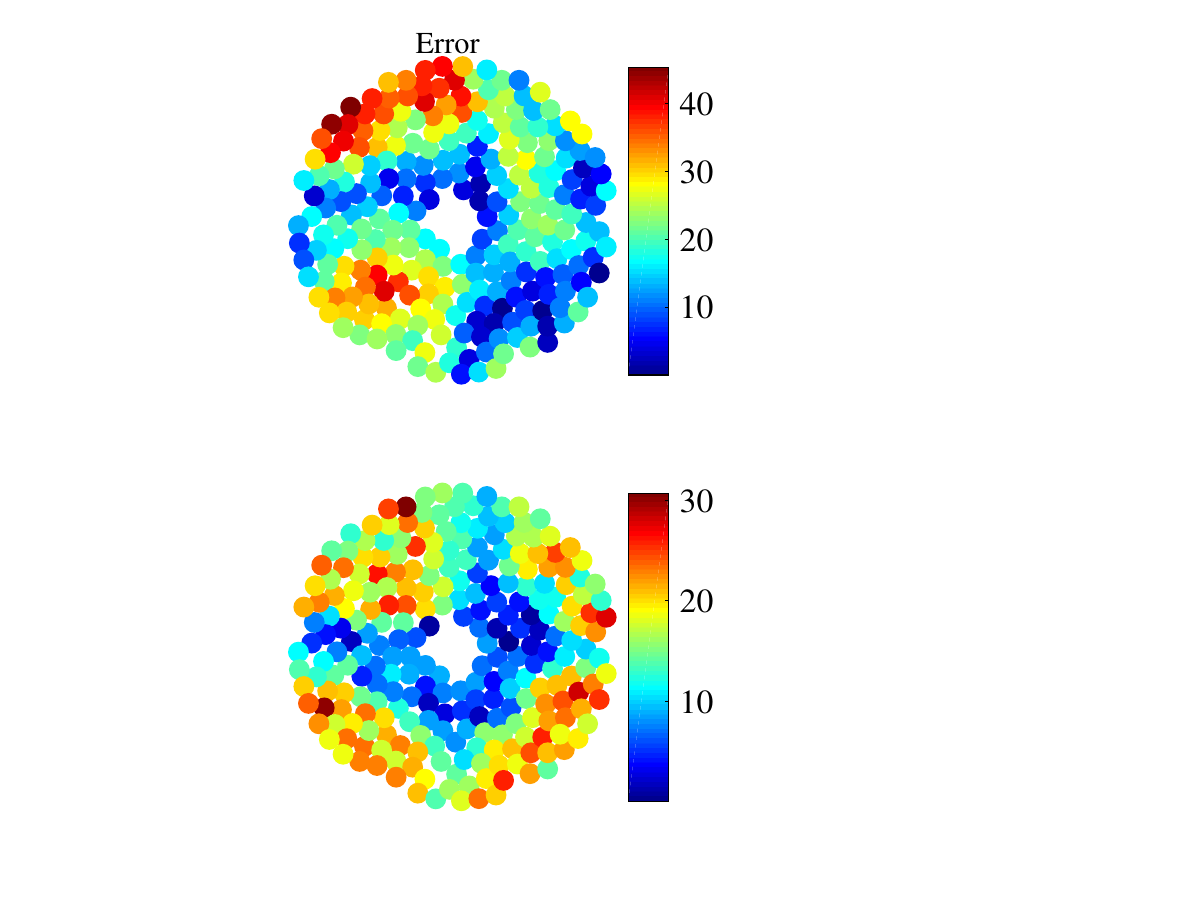}} 
 \caption{}  
  \end{subfigure}
  \begin{subfigure}[b]{0.23\textwidth}
  \includegraphics[width = 1 \textwidth, clip=true, trim = 4.5cm 8cm 7.8cm
  0cm]{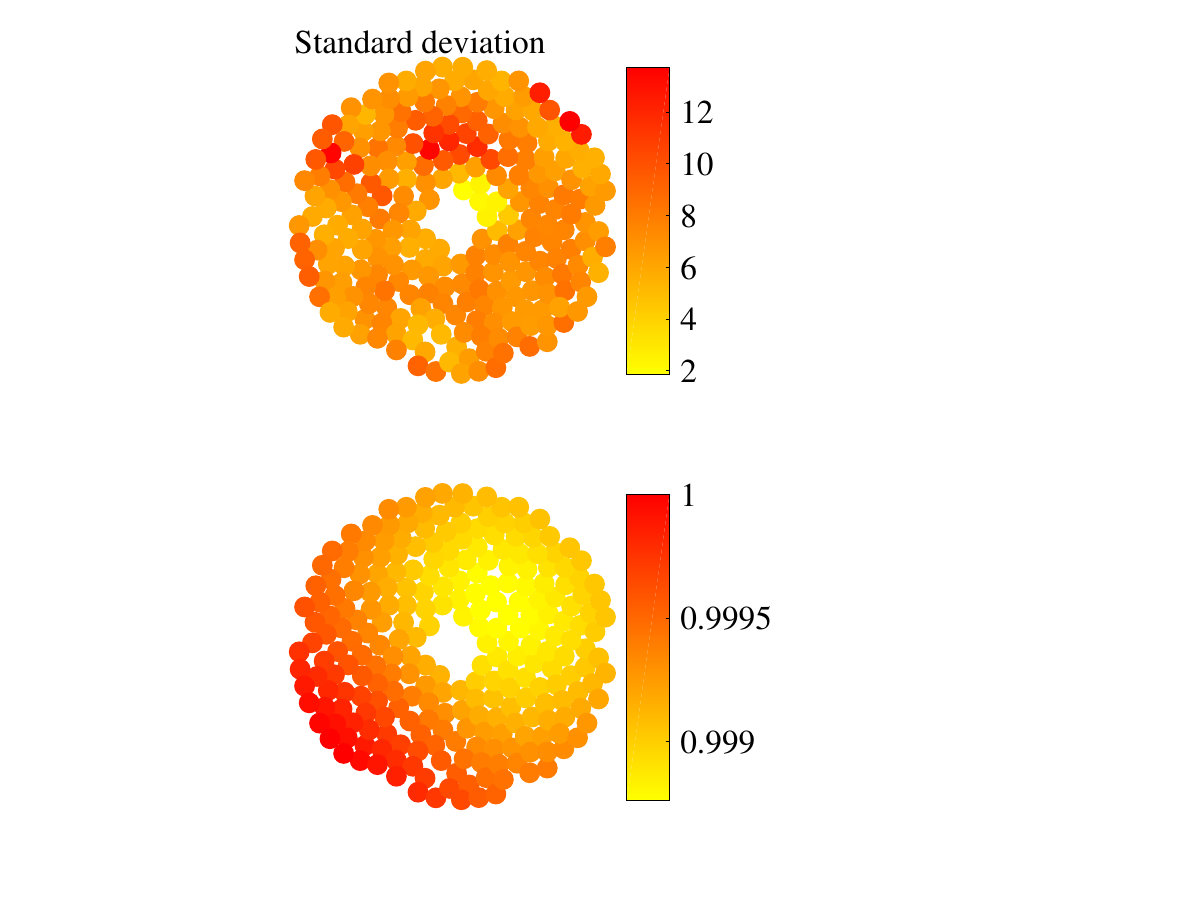} \caption{}  
  \end{subfigure}
\\
  \begin{subfigure}[b]{0.24\textwidth}
{\includegraphics[width = .92 \textwidth, clip=true, trim = 0cm 0cm .5cm
  12cm]{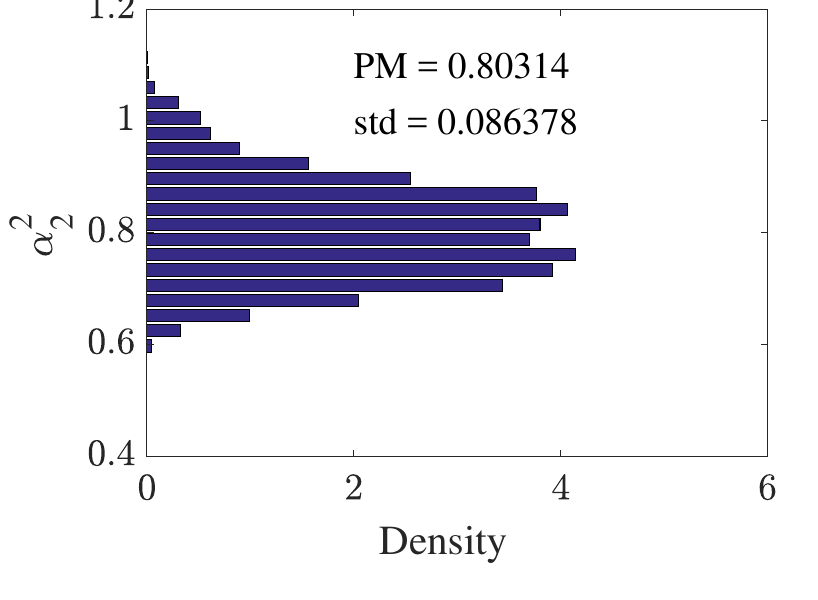}} 
 \caption{}  
  \end{subfigure}
  \begin{subfigure}[b]{0.74\textwidth}
  \begin{overpic}[width = .24 \textwidth, clip=true, trim = 3.5cm 0cm 0cm
  0cm]{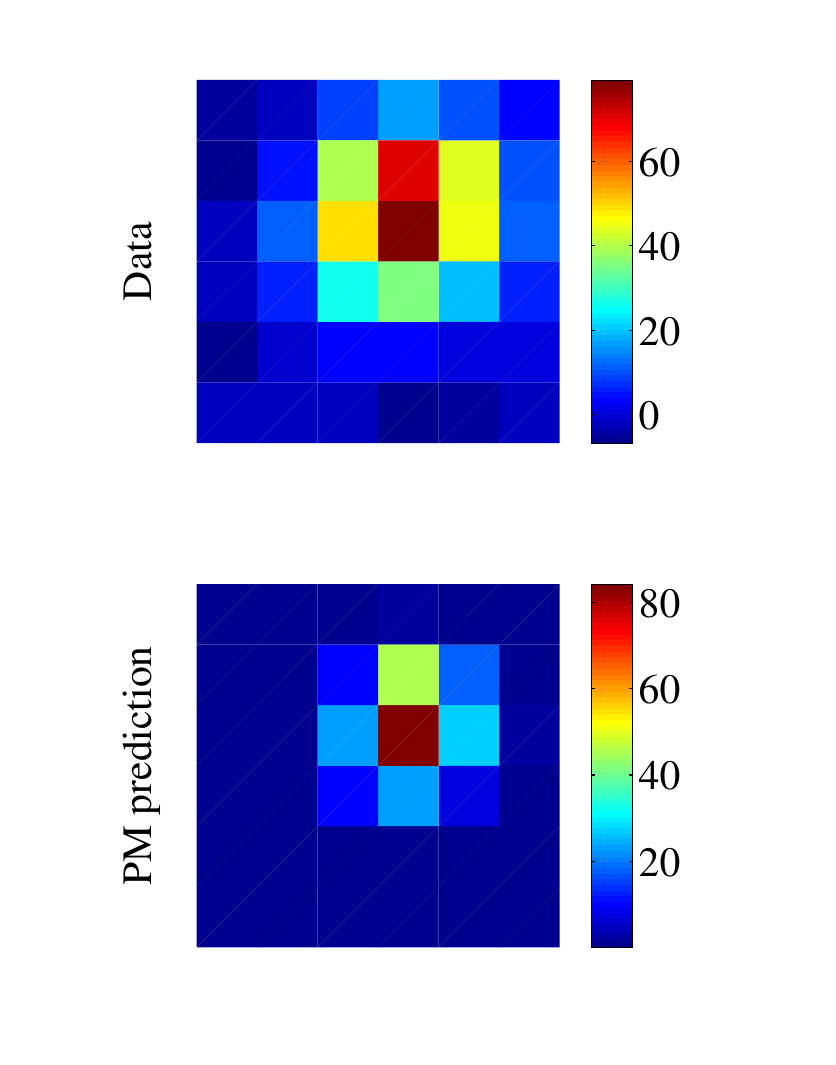}
      \put(-8,55){\rotatebox{90}{\tiny Displacement map}}
      \put(-8,12){\rotatebox{90}{\tiny PM prediction}}
    \end{overpic}
  \includegraphics[width = .32 \textwidth, clip=true, trim = 0cm 0cm 0cm
  0cm]{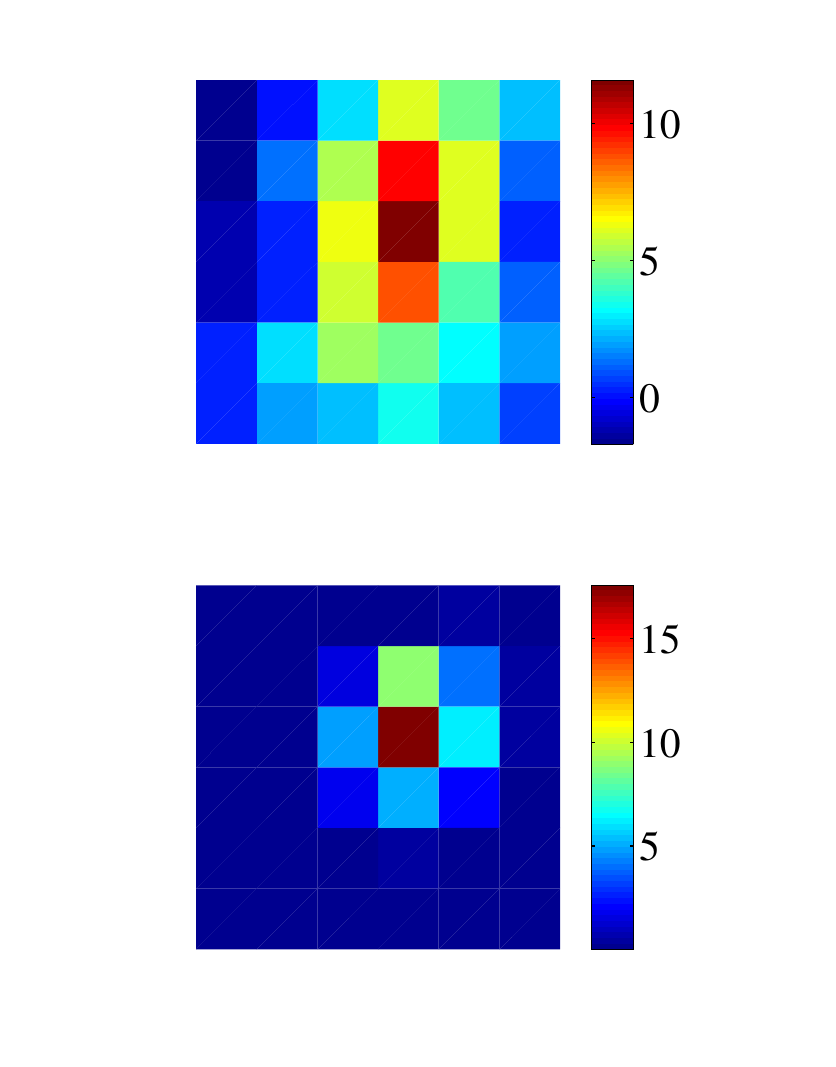}
  \includegraphics[width = .32 \textwidth, clip=true, trim = 0cm 0cm 0cm
  0cm]{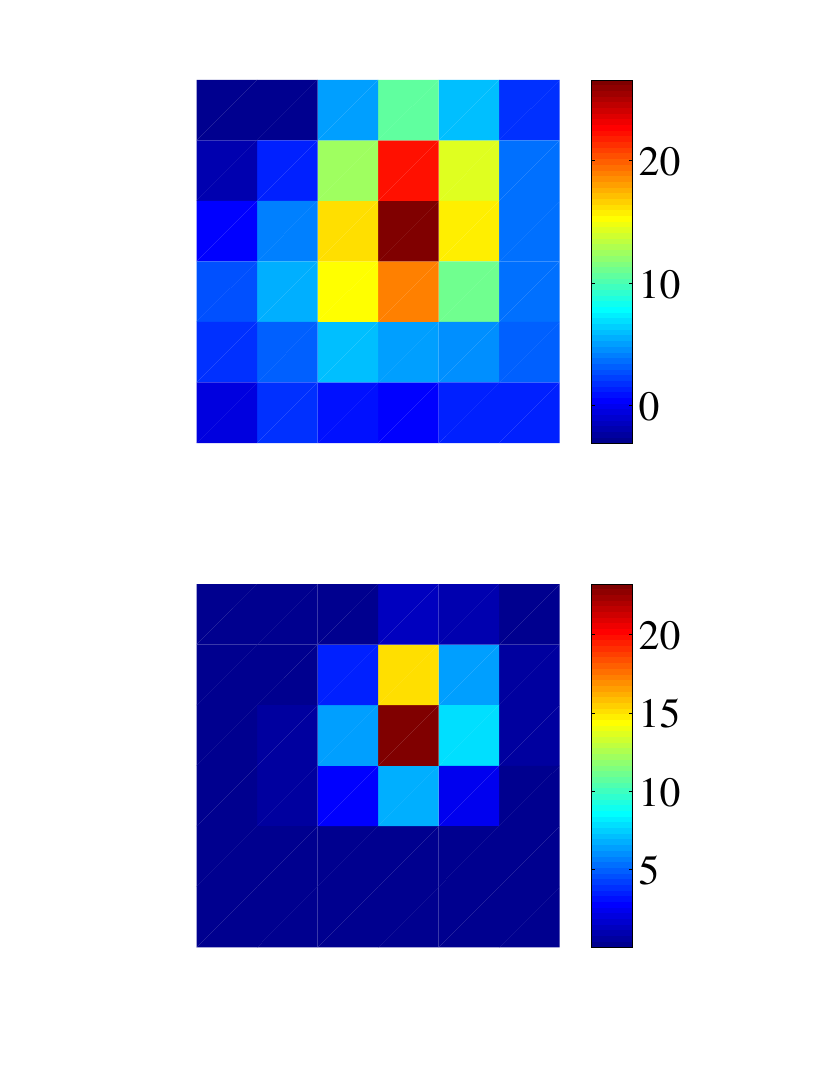}
 \caption{}  
  \end{subfigure}
 \caption{A summary of Bayesian posterior statistics using the
   experimental MR-ARFI
    dataset. (a) The target aberrator imposed on the transducer. (b)
The posterior mean (PM) phase shift and (c) pointwise difference between the
posterior mean and the target aberrator. (d)
The standard deviation of the posterior. 
(e) The marginal distribution of the hyperparameters estimated using
the Markov chain. (f) A comparison between a few frames of the MR-ARFI
displacement map and the prediction of the forward model at the posterior mean.}
  \label{fig:CM-std-reconstruction-experimental}
\end{figure}

Figures 8 to 10 show a similar summary of the results for the
test with MR-ARFI displacement map as was shown for the test with
synthetic displacement maps. Here the results 
were computed using $5 \times 10^5$ samples
from the Markov chain with a burn-in period of $3 \times 10^5$.
\bhedit{Figure \ref{fig:CM-std-reconstruction-experimental}(b) shows the
posterior mean of the phase shift $\pmb{\phi}_{\text{PM}}$ which is once again taken to be a good estimator of the
true phase shift. Figure
\ref{fig:CM-std-reconstruction-experimental}(c)
depicts the pointwise absolute error between $\pmb{\phi}_{\text{PM}}$ and
the target.} Here, the maximum error in the phase is $45$ degrees and
the average error across the elements is $19$ degrees.
The errors here are notably larger as compared to the test with
the synthetic displacement map,
 especially in the case of the attenuation.
This is most likely due to large discrepancies between the forward
model and the measured MR-ARFI data.
 Nevertheless, the overall shape of the
aberrator and range of phase shifts are captured. 

Figure
\ref{fig:CM-std-reconstruction-experimental}(d) shows an estimate of
the standard deviation of the aberrations under $\pi_{\text{post}}$ indicating a
possible error of plus or minus 12 degrees. It should be noted that
this error is still smaller than the true pointwise error and the
algorithm underestimate the true uncertainty of the solution.
Figure
\ref{fig:CM-std-reconstruction-experimental}(e) shows the posterior
marginal on the hyperparameters. As before, the posterior standard
deviation on the hyperparameters has 
been reduced significantly as compared to that of the prior distribution which 
is a sign that these parameters are well identified by the
data. To check whether the prediction of $\mb{a}_{\text{PM}}$ matches
the data, a
 few frames of the
MR-ARFI images are compared to the prediction at the posterior mean in Figure
\ref{fig:CM-std-reconstruction-experimental}(f).
The matching between the data and the prediction was adequate 
at the focal point but deteriorates away from it. Specifically, the
prediction obtained
 the correct range of
variations of each frame and the right location for the point of
maximum intensity. 

For this dataset
$EI(\mb{a}_{\text{PM}}) \approx 42\%$ which predicts a good recovery of
the intensity. However, expected recovery
 $ER(\mb{a}_{\text{PM}}) \approx 3\% $ which is smaller in comparison
 to 
the synthetic test above. This is
expected since the estimate of the aberrator is not as accurate as
before due to discrepancies between the model and the physical data.

\begin{figure}[htp]
  \centering
  \begin{overpic}[width = 1 \textwidth, clip=true, trim = 5cm 8cm 4.5cm
  .5cm]{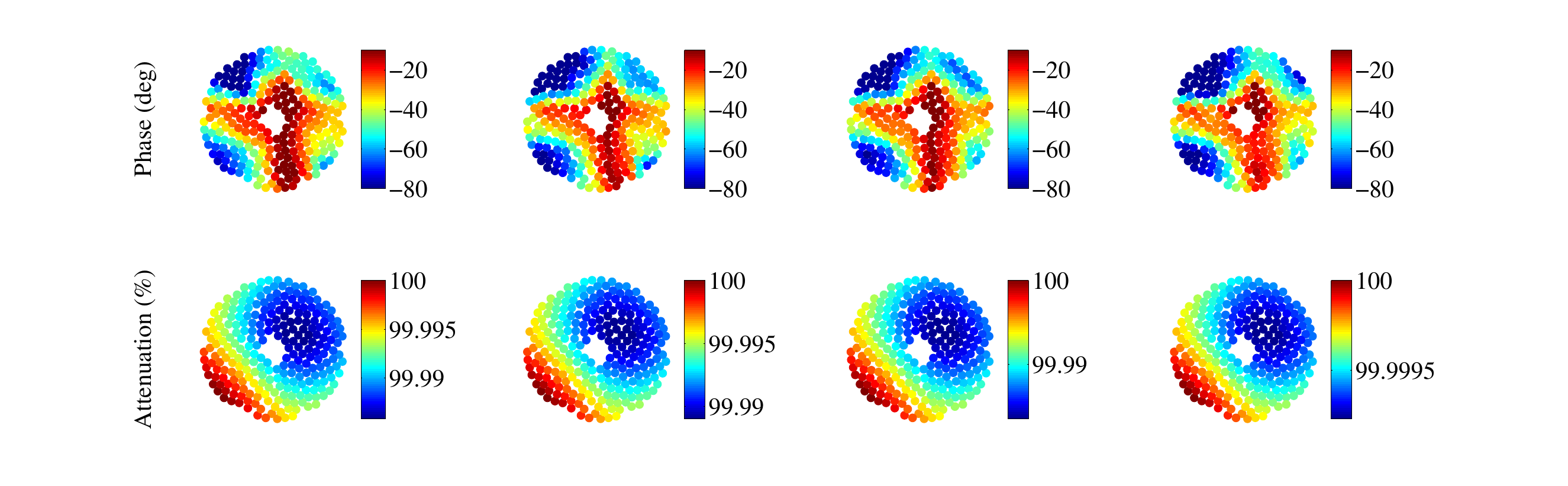}
    \put(0,2){\rotatebox{90}{\tiny Phase shift (deg)}}
  \end{overpic}
  \caption{Four samples from the Markov chain on phase shift after the burn-in period with the experimental MR-ARFI dataset. The samples are $10^5$ steps apart so that they can be treated as independent.}
  \label{fig:experimental-Posterior-samples}
\end{figure}

Finally, a few independent samples from $\pi_{\text{post}}$ are presented in 
Figure \ref{fig:experimental-Posterior-samples}. The samples are
$10^5$ steps apart in order to ensure their independence. This choice was based
on the slowest decaying integrated autocorrelation function in Figure
\ref{fig:MCMC-performance}(b). Once again, comparing the samples to
the posterior mean indicates that the samples are close to the
posterior mean and so $\mb{a}_{\text{PM}}$ is a reasonable point
estimator for the parameter values.

\section{Discussion}\label{sec:discussion}
\subsection{Mathematical framework}
The most notable feature of the forward problem in
\eqref{forward-model-full-dataset} is that 
$\mb{d}$ is a non-linear function of the aberration $\mb{a}$. This is
an attribute of the 
MR-ARFI data which makes the inverse problem more challenging to
solve. In particular, this means that one-shot methods for computing
the minimizers of regularized least squares functionals (such as the
common Tikhonov regularization) are no longer applicable. In this case
one can use numerical optimization algorithms such as Newton's method
or 
the L-BFGS algorithm \cite{vogel, nocedal} to
find a minimizer but this choice requires a modification of the
formulation to get around the non-differentiability of the forward map
with respect to the phase.

\begin{figure}[htp]
  \centering
  \begin{overpic}[width= .25\textwidth, clip = true, trim= 2cm 0cm 2cm 0cm]{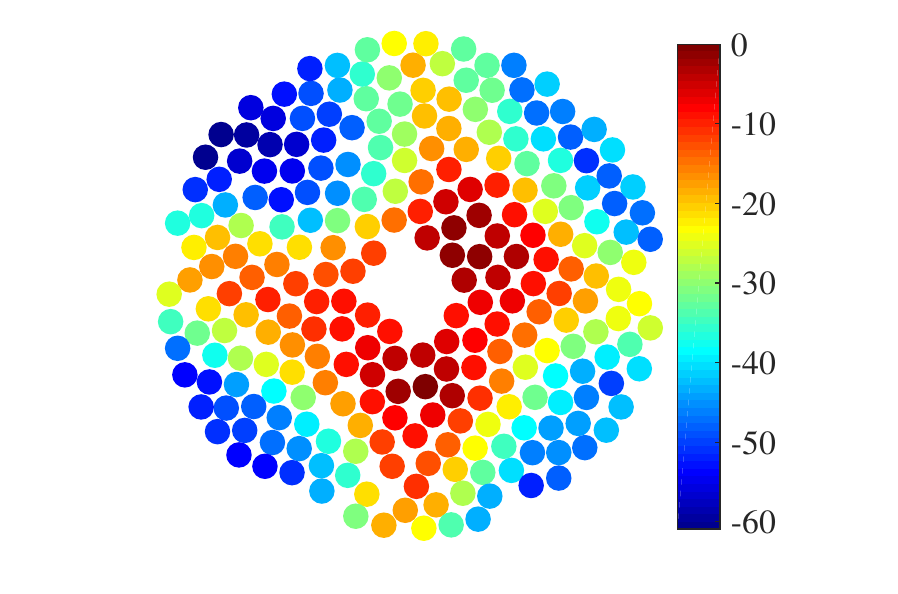}
    \put(35,90){\scriptsize Target}
\end{overpic} 
\begin{overpic}[width= .25\textwidth, clip = true, trim= 2cm 0cm 2cm 0cm ]{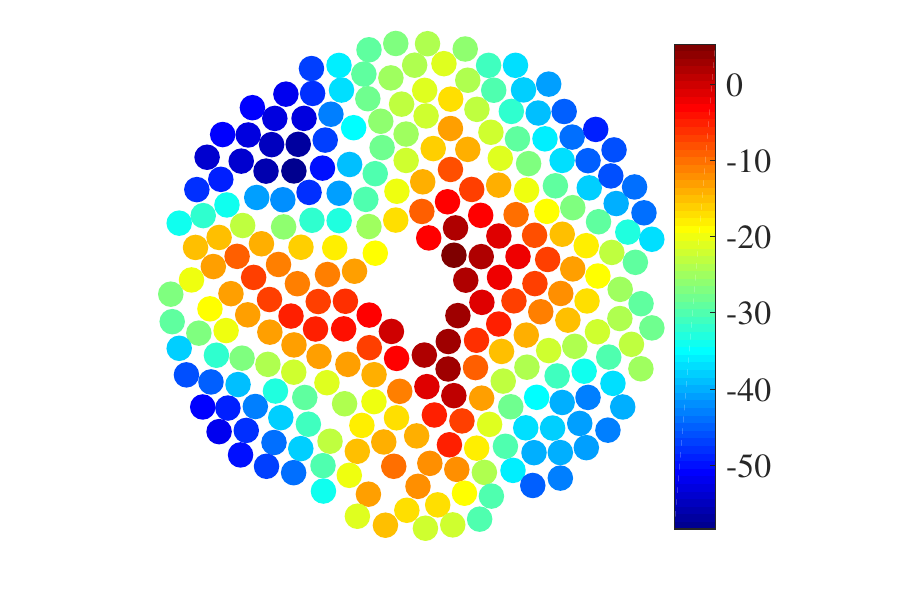}
    \put(25,90){\scriptsize Bayesian PM}
\end{overpic}
\begin{overpic}[width= .25\textwidth, clip = true, trim = 2cm 0cm 2cm 0cm ]{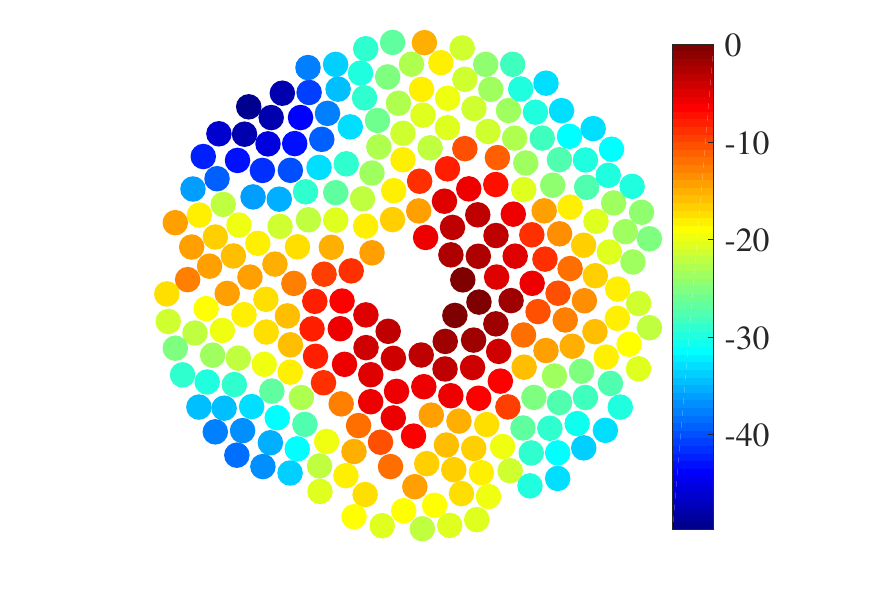}
      \put(15,90){\scriptsize Matrix completion}
\end{overpic}
\\
\hspace{.25\textwidth}
\begin{overpic}[width= .25\textwidth, clip = true, trim= 2cm 0cm 2cm 0cm ]{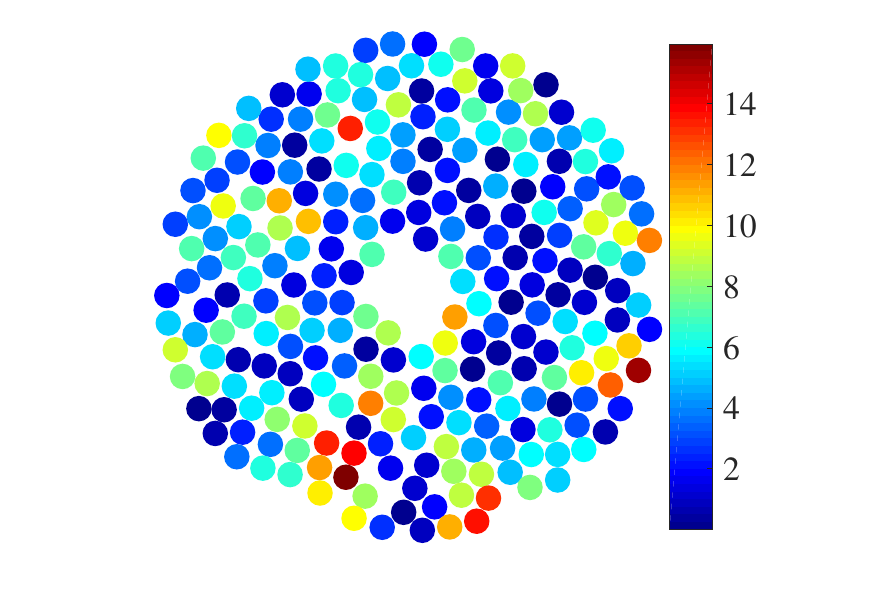}
       \put(15,88){\scriptsize Bayesian PM error}
\end{overpic}
\begin{overpic}[width= .25\textwidth, clip = true, trim = 2cm 0cm 2cm 0cm ]{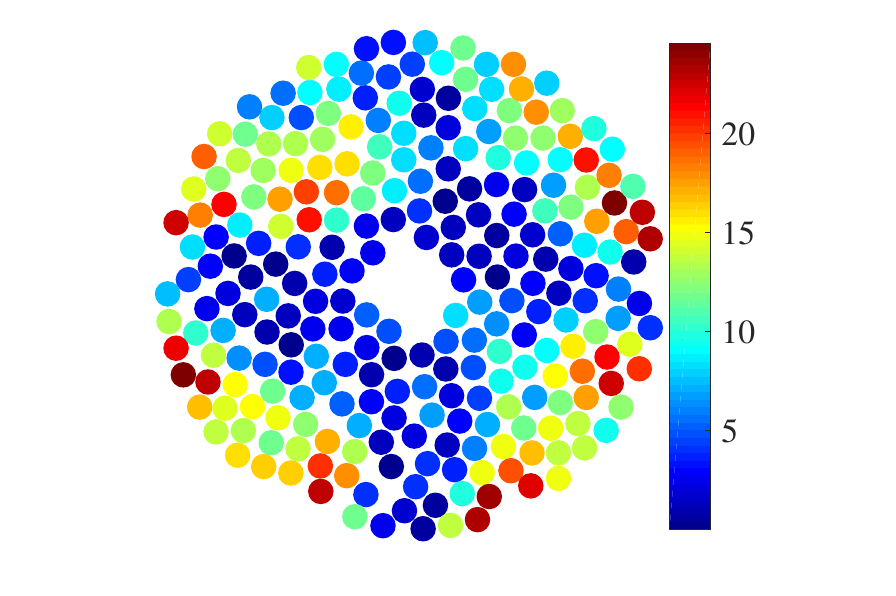}
\put(0,88){\scriptsize Matrix completion error}
\end{overpic}
  \caption{{\color{black}Comparison between phase shift estimators 
obtained using the Bayesian approach and the matrix completion method of 
\cite{candes-phaseretrieval} (top row), along with the corresponding 
absolute error between the target and estimate (bottom row). 
All values are reported in degrees.}}
  \label{fig:matrix-completion-comparison}
\end{figure}

\bhedit{Another effective method for estimating the phase shift, which
  lies within the optimization category, 
 is the matrix completion approach of \cite{candes-phaseretrieval}. 
Matrix completion recasts the phase retrieval problem 
as a penalized least squares problem
 in a high dimensional space with a linear forward map
 \eqref{forward-model-full-dataset}. The resulting solution is a
 low-rank square matrix and the phase shift is given by  
the first eigenvector of that matrix.  
 In Figure~\ref{fig:matrix-completion-comparison} we present a comparison 
between the Bayesian approach and the matrix completion method for the 
synthetic dataset of Section~\ref{sec:synth-disp}, where the matrix completion problem 
was solved
using the TFOCS package \cite{becker2011templates, TFOCS}.  We note
first that matrix completion is more efficient than the Bayesian
approach, since the minimizer can be found in about 15 minutes 
while the Bayesian method requires an hour to obtain an stable estimate of the posterior mean and 
variance (both computations were performed on an
Apple MacBook Pro laptop with a 2.3~GHz dual-core Intel Core i5
processor). However, the Bayesian solution appears to be significantly more 
accurate. Furthermore, a comparison to Figure~\ref{fig:artificial-CM-std-reconstruction}(d) shows that the uncertainty estimate obtained using the Bayesian method 
serves as a certificate that indicates the order of the error in the Bayesian estimate; 
such an uncertainty estimate is not available for  matrix completion. Finally, 
we tuned the regularization parameter in the matrix completion problem to obtain 
a good estimate of the phase shift. The Bayesian method on the other hand 
estimates the hyperparameters automatically.}

From a practical perspective,
our formulation of the forward
and inverse problems  offer complete freedom in the choice of the number of
sonications tests $J$, the size of the MRI images $M$, the size of the 
parameter space for the aberrations $G$ and the number of elements
$N$. It is often preferable to take $J$ to be as small as possible to reduce
the time required to collect the MRI data. It is also
desirable to choose $G$ to be small in order to reduce the
computational cost of the algorithms. Taking $M$ to be large means
that each MRI image will have more information. However, the
signal to noise ratio drops rapidly for voxels that are far from the
focal point. This is the  reason why the dataset in the MR-ARFI
experiments uses a  smaller field of view compared to the
artificial example. Then the choice of each one of these parameters requires further study in future.

\subsection{Performance of the MCMC algorithm}
The pCN algorithm is a modification of the random walk
Metropolis Hastings algorithm that is well defined on a function space
\cite{stuart-mcmc}. 
It tends
to generate samples that are highly correlated and so it explores the
posterior distribution slowly. However, the pCN update does not require derivative
information and therefore can be used to sample from
non-differentiable densities. This also means that each step of pCN is
relatively inexpensive.
 The MALA algorithm utilizes an optimal proposal step that results in 
less correlated samples
\cite{casella, stuart-mcmc} which makes the algorithm better
at exploring the posterior distribution $\pi_{\text{post}}$ but each step of the algorithm
has a higher computational cost. This cost becomes significant  when 
a large dataset is at hand, because this increases the cost of
gradient computations. 

\begin{figure}[htp]
  \centering
  \begin{subfigure}[b]{0.49 \textwidth}
 \includegraphics[width = .32 \textwidth, clip=true, trim = 0.0cm 0cm 0.5cm
  0.2cm]{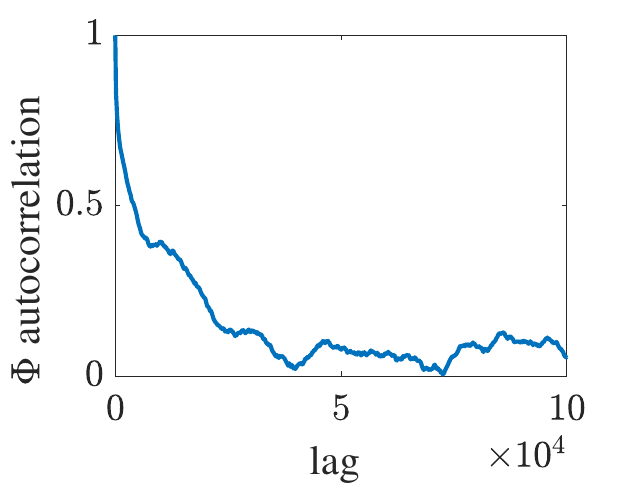}  
      \includegraphics[width = .65 \textwidth, clip=true, trim = 0.cm 0cm 1.5cm
  0.2cm]{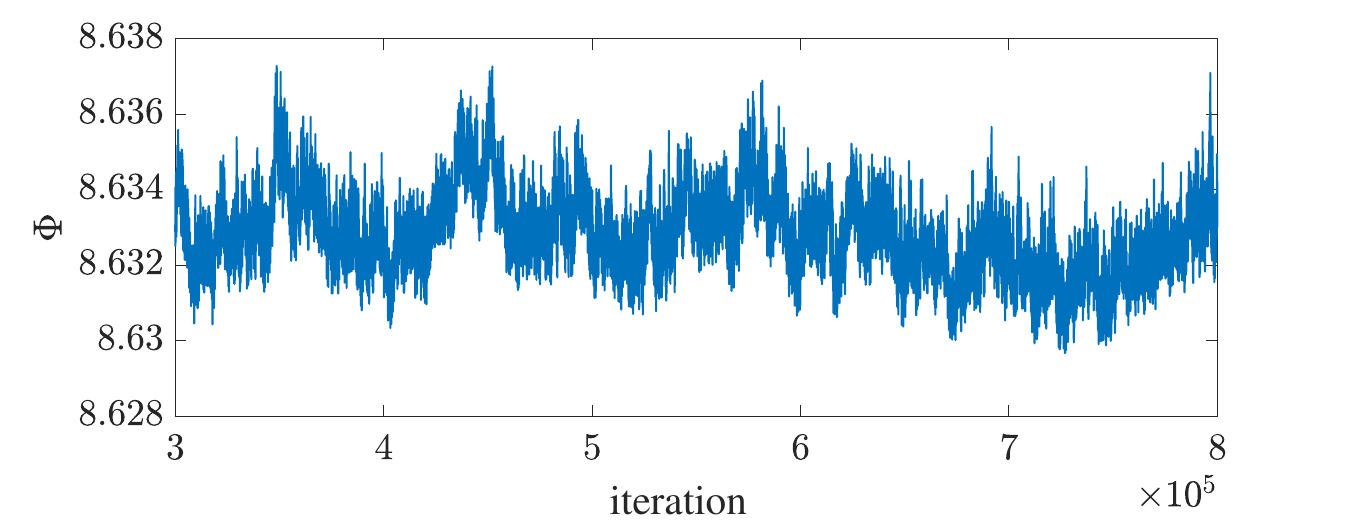} \\

      \includegraphics[width = .33 \textwidth, clip=true, trim = 0.0cm 0cm 0.5cm
  0.2cm]{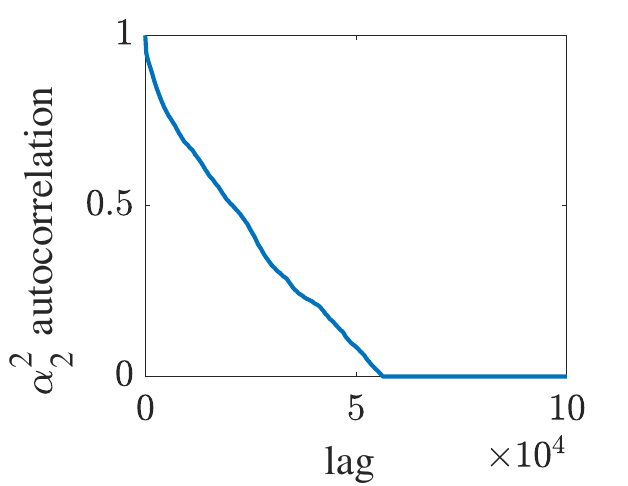} 
      \includegraphics[width = .65 \textwidth, clip=true, trim = 0.cm 0cm 1.5cm
  0.2cm]{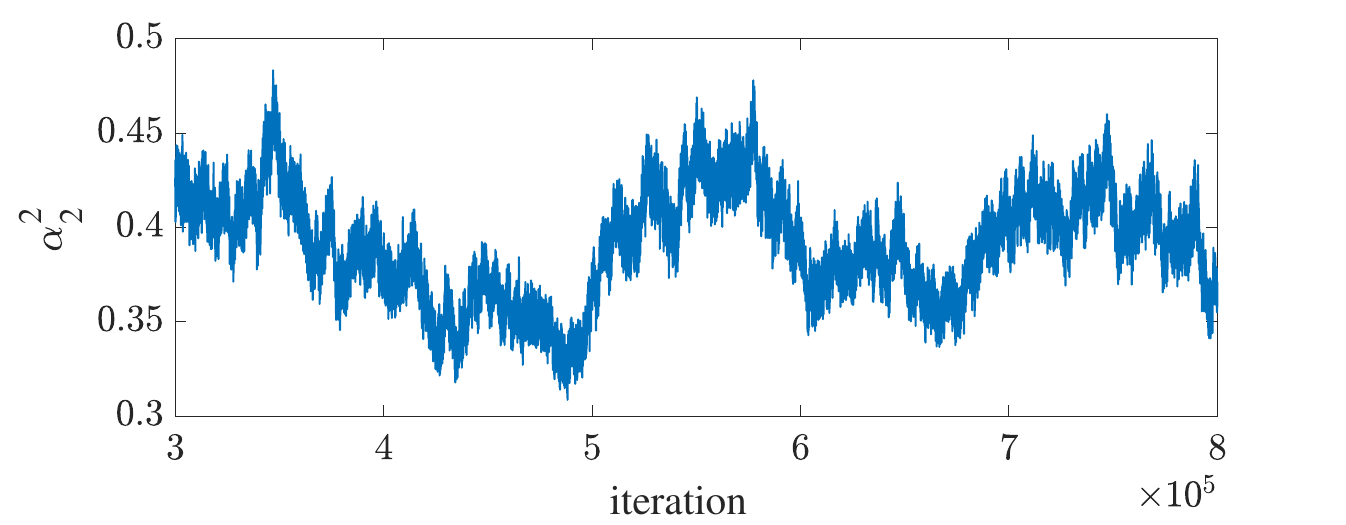} \\  
      \includegraphics[width = .33\textwidth, clip=true, trim = 0.0cm 0cm 0.5cm
  0.2cm]{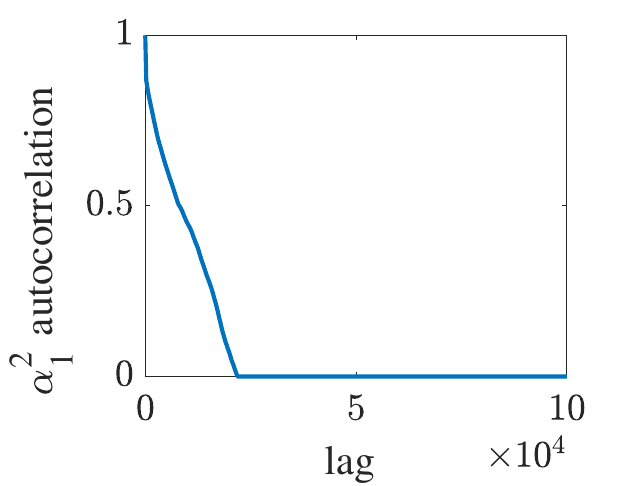} 
      \includegraphics[width = .65 \textwidth, clip=true, trim = 0.cm 0cm 1.5cm
  0.2cm]{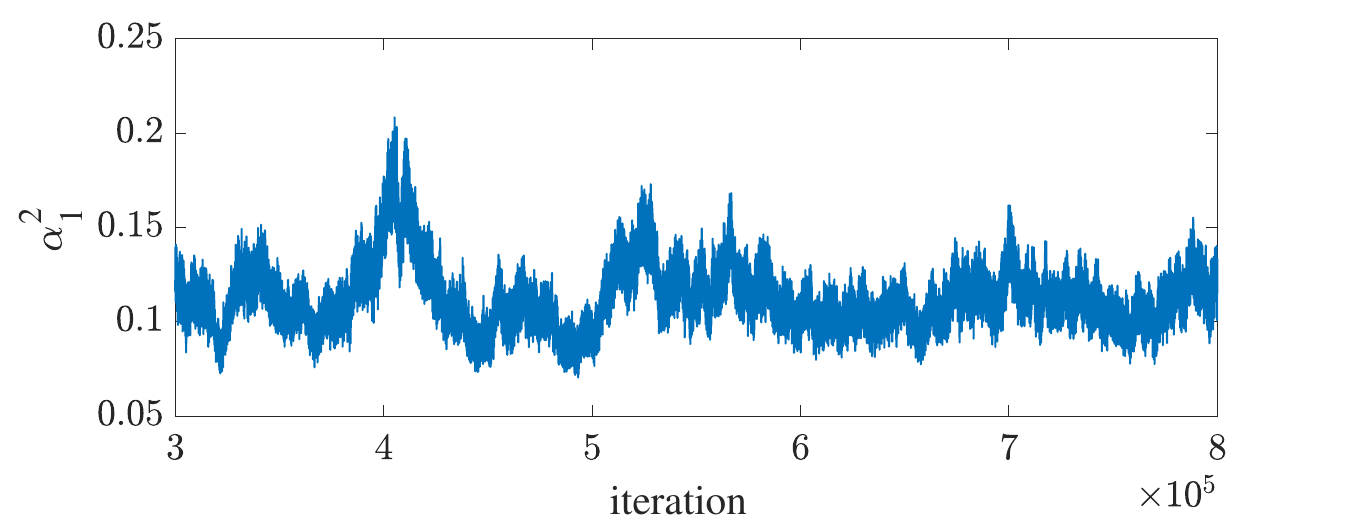}
\caption{}
  \end{subfigure}
\begin{subfigure}[b]{0.5 \textwidth}
  \raisebox{0.05\textwidth}{\includegraphics[width = .32\textwidth, clip=true, trim = 0.0cm 0cm 0.5cm
  0.2cm]{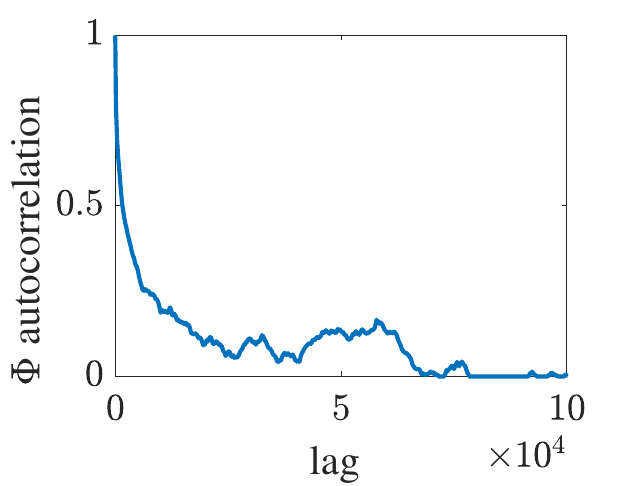}
      \includegraphics[width = .65\textwidth, clip=true, trim = 0.cm 0cm 1.5cm
  0.2cm]{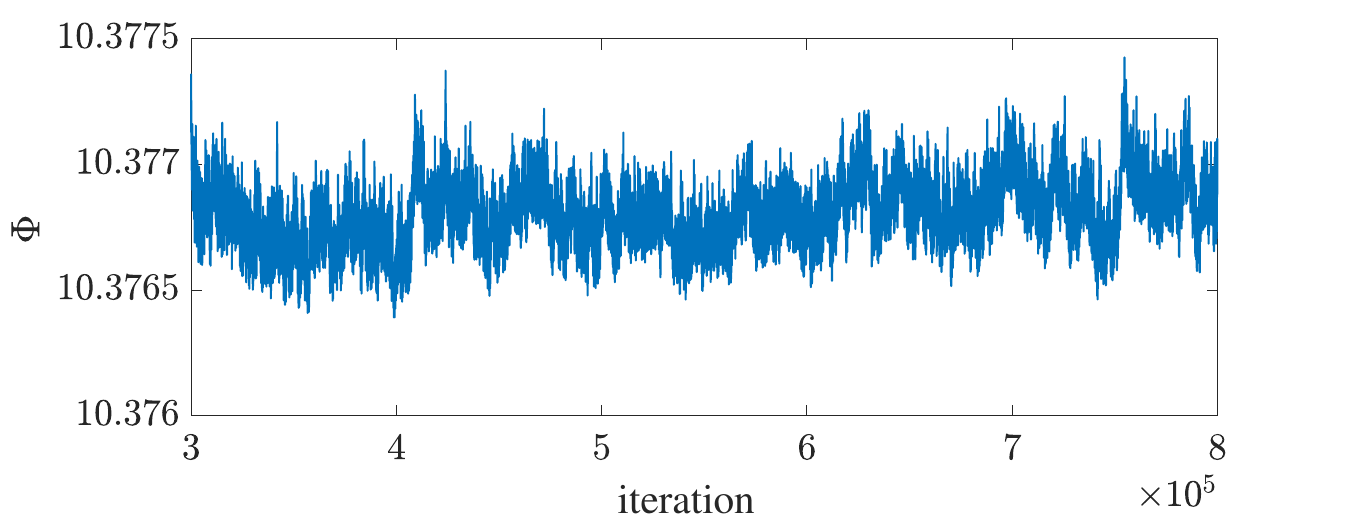}} \\
      \raisebox{.26\textwidth}{\includegraphics[width = .33 \textwidth, clip=true, trim = 0.0cm 0cm 0.5cm
  0.2cm]{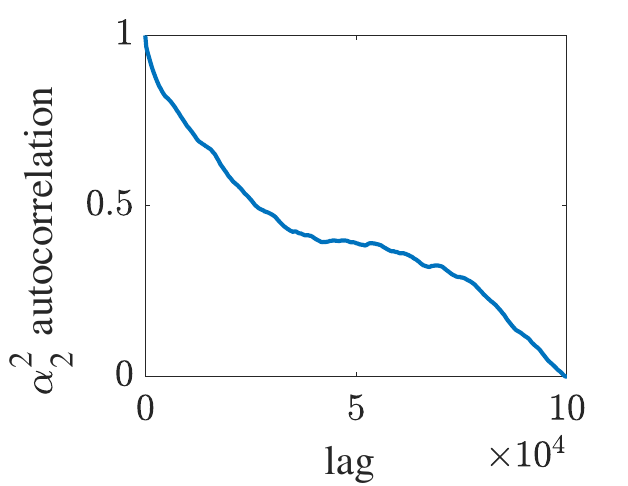} 
      \includegraphics[width = .65 \textwidth, clip=true, trim = 0.cm 0cm 1.5cm
  0.2cm]{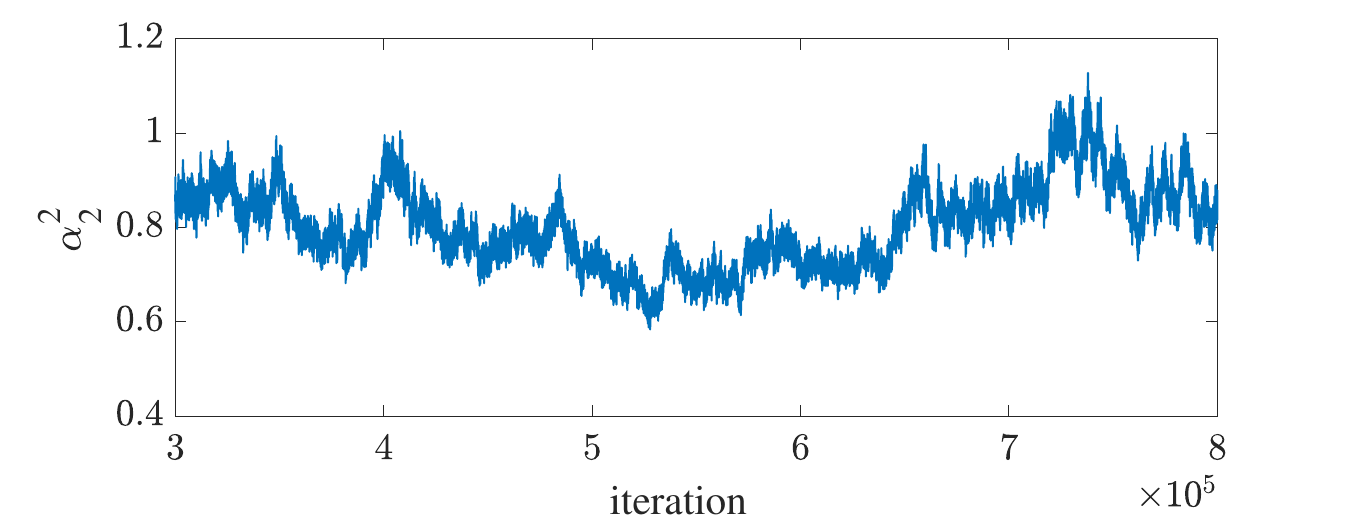}} 
\caption{}
  \end{subfigure}
  \caption{Autocorrelation and trace plots of the likelihood potential
and
    hyperparameters using (a) the synthetic dataset and (b) the
    experimental MR-ARFI displacement map.}
  \label{fig:MCMC-performance}
\end{figure}

The difference between the two steps of the algorithm is apparent in
Figure \ref{fig:MCMC-performance}(a) where the integrated autocorrelation functions and
trace plots are presented for the likelihood potential $\Phi$ and the
\bhedit{squared hyperparameters
$\alpha_1^2$ and $\alpha_2^2$ in the test with the synthetic dataset.
The integrated autocorrelation of $\alpha_2^2$ decays
slower than that of $\alpha_1^2$. This is likely because samples
from the latter are generated using the MALA updates. The $\alpha_2^2$
chain is slow because in this case the data is fairly informative in
the direction of phase shift,
meaning that the posterior distribution $\pi_{\text{post}}$ is
dominated by the likelihood rather than the prior
distribution. Looking at the trace plots in Figure
{\ref{fig:MCMC-performance}}, it is interesting to note that the $\Phi$ chain 
demonstrates better mixing in comparison to the chains for $\alpha_1^2$
and $\alpha_2^2$. We observe a similar behavior in the case of the physical dataset.}
  

The results of this article were obtained using an implementation of
the MwG algorithm for
estimation of the aberrator in MATLAB on a personal laptop (MacBook Pro with a 3GHz
Intel Core i7 processor and 8 GB or memory) in less than two hours.
 This processing time may not be ideal for practical
settings but it can be effectively reduced by 
implementation of the algorithm in more efficient languages
and applying techniques such as parallel
tempering and population MCMC \cite{liu-mc} to further
reduce the required sample size needed to compute the expectations.

\subsection{Quality of the estimates with synthetic and
  physical datasets}

In Section~\ref{sec:synth-disp} the results with a synthetic dataset were presented.
 This example is viewed as an idealized setting
where there is little discrepancy between the process for generation
of  the data
and the forward model. The  reconstructions
were adequate even in the presence of
large errors, the posterior
mean $\mb{a}_{\text{PM}}$ of the phase (Figure \ref{fig:artificial-CM-std-reconstruction}(b)) was 
 close to the actual value of the 
aberrations (the average pointwise error of the phase estimate was
4.5 degrees amounting to $15\%$ relative error). Furthermore, the estimated standard deviation appeared to
be a good estimate of the expected errors in the
reconstruction. However, in the case of the attenuation,
the posterior mean $\mb{a}_{\text{PM}}$ was not as accurate
as the phase shift even in this idealized setting (compare Figure
\ref{fig:artificial-CM-std-reconstruction}(a) and (b)) . This was due
to the fact that the displacement data was
not sensitive to relatively small changes in the
attenuation.

\bhedit{In Section~\ref{sec:results} we used the expected improvement $EI$ and expected 
recovery $ER$ functionals as a measure of the quality of reconstructions. Recall that 
$EI$ is a measure of improvement in power efficiency  while $ER$ is a measure 
of improvement in required dosage. In the synthetic and physical tests we recovered 
3 to 5\% of the total intensity after refocusing. This was partly due to the 
fact that we used a weak aberrator that reduced the beam intensity by
$7\%$, so that the beam 
was already at $93\%$ intensity without any focusing. Of course this makes our aberrator 
harder to estimate as compared to a stronger aberrator. However, even in
this setting, a $3$ to $5\%$ improvement in beam intensity results in a similar
gain in power efficiency but has a more significant impact in dosage efficiency 
due to the fact that tissue damage grows exponentially  \cite{sapareto1984thermal} as intensity 
(and in turn temperature) increases at the focal point.}

Finally, we note that the computed free-field matrix of the transducer and the calibration
step have a significant impact on
the quality of the reconstructions. In the test with experimental
MR-ARFI displacement map of
Section~\ref{methods:physical-test} the free-field was estimated using $160$ sonication 
tests with ten steps in phase for each virtual elements. Using subsets
of this data resulted in a less accurate free-field matrix which in
turn resulted in very different reconstructions. Then one might prefer
to use a large dataset in order to obtain an accurate estimate of the
free-field matrix which can be used on a smaller reconstruction
dataset later on. The fact that calibration can be done offline 
makes this a viable approach in practical settings.


\subsection{Future research}
The framework presented in this article can be extended in multiple directions
in order to improve the quality of the reconstructions which, in turn,
will lead to better focusing of the beam.
Obtaining an
accurate estimate of the empirical free-field with the minimum number
of sonication tests is a crucial task. The 
estimates of the aberrations become more sensitive to the free-field matrix 
as the dataset becomes smaller. Therefore, a good estimate
of the free-field matrix is needed in order to further reduce the 
number sonication tests that are performed under {\it in vivo} or clinic conditions.

Another promising direction for future research is improving the
sampling algorithm. MCMC algorithms often have a difficult time 
traversing high dimensional distributions. In such
cases, strategies such as population or adaptive MCMC or parallel
tempering \cite{liu-mc} can be used to improve the statistical
performance of the chain. Alternatively, one can also improve the 
algorithm by changing the forward model so that it is differentiable
in the phase. This would allow the use of the MALA update on the
entire posterior which would greatly improve the performance. 

On the topic of MR-ARFI experiments, one can explore multiple
directions for improving the quality of the dataset. An interesting
question is the interplay between voxel size, measurement noise and 
acquisition duration. Smaller voxels give a better estimate of the
free-field matrix and the aberration but they are associated with more noise. This, in
turn, requires smaller phase steps and longer acquisition time or perhaps
more averaging steps per sonication test. Therefore, finding the optimal
parameters for generating the dataset remains a challenge in practice.



\section*{Acknowledgements}
The authors would like to thank Profs. Nilima Nigam and Chris Budd for
fruitful discussions. BH and CM are thankful to the Fields Institute and the
organizers of the Fields-Mprime Industrial Problem Solving Workshop
during the 
August of 2014,
where their collaboration was initiated. Finally, this work was
supported in part by the Natural Sciences and Engineering
Research Council of Canada,
the Brain Canada Multi-Investigator
Research Initiative and the Focused Ultrasound Foundation.

\bibliographystyle{abbrv}   
\bibliography{ref}

\begin{thebibliography}{10}

\bibitem{agapiou}
S.~Agapiou, J.~M. Bardsley, O.~Papaspiliopoulos, and A.~M. Stuart.
\newblock Analysis of the {G}ibbs sampler for hierarchical inverse problems.
\newblock {\em SIAM/ASA Journal on Uncertainty Quantification}, 2(1):511--544,
  2014.

\bibitem{arridge-dot}
S.~R. Arridge, J.~P. Kaipio, V.~Kolehmainen, M.~Schweiger, E.~Somersalo,
  T.~Tarvainen, and M.~Vauhkonen.
\newblock Approximation errors and model reduction with an application in
  optical diffusion tomography.
\newblock {\em Inverse Problems}, 22:175--195, 2006.

\bibitem{aubry-CT}
J.~F. Aubry, M.~Tanter, M.~Pernot, J.~L. Thomas, and M.~Fink.
\newblock Experimental demonstration of noninvasive trans-skull adaptive
  focusing based on prior computed tomography scans.
\newblock {\em Journal of the Acoustical Society of America}, 113(1):84--93,
  2003.

\bibitem{becker2011templates}
S.~R. Becker, E.~J. Cand{\`e}s, and M.~C. Grant.
\newblock Templates for convex cone problems with applications to sparse signal
  recovery.
\newblock {\em Mathematical Programming Computation}, 3(3):165, 2011.

\bibitem{TFOCS}
S.~R. Becker, E.~J. Cand{\`e}s, and M.~C. Grant.
\newblock {TFOCS}: Templates for first-order conic solvers.
\newblock {http://cvxr.com/tfocs/}, May 2018.

\bibitem{calvetti}
D.~Calvetti and E.~Somersalo.
\newblock {\em {An Introduction to Bayesian Scientific Computing: Ten Lectures
  on Subjective Computing}}, volume~2.
\newblock Springer Science and Business Media, New York, 2007.

\bibitem{candes-phaseretrieval}
E.~J. Cand{\`e}s, Y.~C. Eldar, T.~Strohmer, and V.~Voroninski.
\newblock Phase retrieval via matrix completion.
\newblock {\em SIAM Review}, 57(2):225--251, 2015.

\bibitem{chen2010optimization}
J.~Chen, R.~Watkins, and K.~B. Pauly.
\newblock Optimization of encoding gradients for {MR-ARFI}.
\newblock {\em Magnetic Resonance in Medicine}, 63(4):1050--1058, 2010.

\bibitem{elodie}
E.~Constanciel~Colas, A.~C. Waspe, C.~Mougenot, T.~Looi, S.~Pichardo, and J.~M.
  Drake.
\newblock Mapping of insertion losses and time-of-flight delays of pediatric
  skulls using a clinical {MR}-guided high intensity focused ultrasound system.
\newblock In {\em International Society for Therapeutic Ultrasound}, Utrecht,
  Netherlands, April 2015.

\bibitem{stuart-mcmc}
S.~L. Cotter, G.~O. Roberts, A.~M. Stuart, and D.~White.
\newblock {MCMC} methods for functions: modifying old algorithms to make them
  faster.
\newblock {\em Statistical Science}, 28(3):424--446, 2013.

\bibitem{crouzet}
S.~Crouzet, F.~J. Murat, G.~Pasticier, P.~Cassier, J.~Y. Chapelon, and
  A.~Gelet.
\newblock High intensity focused ultrasound ({HIFU}) for prostate cancer:
  current clinical status, outcomes and future perspectives.
\newblock {\em International Journal of Hyperthermia}, 26(8):796--803, 2010.

\bibitem{fox-geothermal}
T.~Cui, C.~Fox, and M.~J. O'Sullivan.
\newblock Bayesian calibration of a large-scale geothermal reservoir model by a
  new adaptive delayed acceptance {M}etropolis {H}astings algorithm.
\newblock {\em Water Resources Research}, 47(10):{W10521}, 2011.

\bibitem{edwards-supernovae}
M.~C. Edwards, R.~Meyer, and N.~Christensen.
\newblock Bayesian parameter estimation of core collapse supernovae using
  gravitational wave simulations.
\newblock {\em Inverse Problems}, 30(11):114008, 2014.

\bibitem{elias2013pilot}
W.~J. Elias, D.~Huss, T.~Voss, J.~Loomba, M.~Khaled, E.~Zadicario, R.~C.
  Frysinger, S.~A. Sperling, S.~Wylie, S.~J. Monteith, et~al.
\newblock A pilot study of focused ultrasound thalamotomy for essential tremor.
\newblock {\em New England Journal of Medicine}, 369(7):640--648, 2013.

\bibitem{fennessy}
F.~M. Fennessy and C.~M. Tempany.
\newblock A review of magnetic resonance imaging-guided focused ultrasound
  surgery of uterine fibroids.
\newblock {\em Topics in Magnetic Resonance Imaging}, 17(3):173--179, 2006.

\bibitem{fienup2013phase}
J.~R. Fienup.
\newblock Phase retrieval algorithms: a personal tour.
\newblock {\em Applied Optics}, 52(1):45--56, 2013.

\bibitem{herbert-energy}
E.~Herbert, M.~Pernot, G.~Montaldo, M.~Fink, and M.~Tanter.
\newblock Energy-based adaptive focusing of waves: application to noninvasive
  aberration correction of ultrasonic wavefields.
\newblock {\em IEEE Transactions on Ultrasonics, Ferroelectrics and Frequency
  Control}, 56(11):2388--2399, 2009.

\bibitem{hynynen}
K.~Hynynen and J.~Sun.
\newblock Trans-skull ultrasound therapy: The feasibility of using
  image-derived skull thickness information to correct the phase distortion.
\newblock {\em IEEE Transactions on Ultrasonics, Ferroelectrics, and Frequency
  Control}, 46(3):752--755, 1999.

\bibitem{Iglesias-subsurface}
M.~A. Iglesias, K.~Lin, and A.~M. Stuart.
\newblock Well-posed {B}ayesian geometric inverse problems arising in
  subsurface flow.
\newblock {\em Inverse Problems}, 30(11):114001, 2014.

\bibitem{ikink}
M.~E. Ikink, M.~J. Voogt, H.~M. Verkooijen, P.~N.~M. Lohle, K.~J. Schweitzer,
  A.~Franx, P.~T.~M. Willem, L.~W. Bartels, and M.~A. A.~J. {van den Bosch}.
\newblock Mid-term clinical efficacy of a volumetric magnetic resonance-guided
  high-intensity focused ultrasound technique for treatment of symptomatic
  uterine fibroids.
\newblock {\em European Radiology}, 23(11):3054--3061, 2013.

\bibitem{illing-safety}
R.~O. Illing, J.~E. Kennedy, F.~Wu, G.~R. {Ter Haar}, A.~S. Protheroe, P.~J.
  Friend, F.~V. Gleeson, D.~W. Cranston, R.~R. Phillips, and M.~R. Middleton.
\newblock The safety and feasibility of extracorporeal high-intensity focused
  ultrasound ({HIFU}) for the treatment of liver and kidney tumours in a
  western population.
\newblock {\em British Journal of Cancer}, 93(8):890--895, 2005.

\bibitem{jeanmonod}
D.~Jeanmonod, B.~Werner, A.~Morel, L.~Michels, E.~Zadicario, G.~Schiff, and
  E.~Martin.
\newblock Transcranial magnetic resonance imaging--guided focused ultrasound:
  noninvasive central lateral thalamotomy for chronic neuropathic pain.
\newblock {\em Neurosurgical Focus}, 32(1):E1, 2012.

\bibitem{somersalo}
J.~Kaipio and E.~Somersalo.
\newblock {\em {S}tatistical and {C}omputational {I}nverse {P}roblems}.
\newblock Springer Science and Business Media, New York, 2005.

\bibitem{kaipio-EIT}
J.~P. Kaipio, V.~Kolehmainen, E.~Somersalo, and M.~Vauhkonen.
\newblock Statistical inversion and {Monte Carlo} sampling methods in
  electrical impedance tomography.
\newblock {\em Inverse Problems}, 16(5):1487--1522, 2000.

\bibitem{kaye}
E.~A. Kaye, Y.~Hertzberg, M.~Marx, B.~Werner, G.~Navon, M.~Levoy, and K.~B.
  Pauly.
\newblock Application of {Z}ernike polynomials towards accelerated adaptive
  focusing of transcranial high intensity focused ultrasound.
\newblock {\em Medical Physics}, 39(10):6254--6263, 2012.

\bibitem{larrat}
B.~Larrat, M.~Pernot, G.~Montaldo, M.~Fink, and M.~Tanter.
\newblock {MR}-guided adaptive focusing of ultrasound.
\newblock {\em IEEE Transactions on Ultrasonics, Ferroelectrics, and Frequency
  Control}, 57(8):1734--1747, 2010.

\bibitem{liberman}
B.~Liberman, D.~Gianfelice, Y.~Inbar, A.~Beck, T.~Rabin, N.~Shabshin,
  G.~Chander, S.~Hengst, R.~Pfeffer, A.~Chechick, et~al.
\newblock Pain palliation in patients with bone metastases using {MR}-guided
  focused ultrasound surgery: a multicenter study.
\newblock {\em Annals of Surgical Oncology}, 16(1):140--146, 2009.

\bibitem{lipsman}
N.~Lipsman, M.~L. Schwartz, Y.~Huang, L.~Lee, T.~Sankar, M.~Chapman,
  K.~Hynynen, and A.~M. Lozano.
\newblock {MR}-guided focused ultrasound thalamotomy for essential tremor: a
  proof-of-concept study.
\newblock {\em The Lancet Neurology}, 12(5):462--468, 2013.

\bibitem{liu-mc}
J.~S. Liu.
\newblock {\em {Monte Carlo Strategies in Scientific Computing}}.
\newblock {Springer Series in Statistics}. Springer, New York, 2008.

\bibitem{liu-hifu}
N.~Liu, A.~Liutkus, J.-F. Aubry, L.~Marsac, M.~Tanter, and L.~Daudet.
\newblock Random calibration for accelerating {MR-ARFI} guided ultrasonic
  focusing in transcranial therapy.
\newblock {\em Physics in {M}edicine and {B}iology}, 60(3):1069, 2015.

\bibitem{marchesini2007invited}
S.~Marchesini.
\newblock Invited article: A unified evaluation of iterative projection
  algorithms for phase retrieval.
\newblock {\em Review of Scientific Instruments}, 78(1):011301, 2007.

\bibitem{marchesini2007phase}
S.~Marchesini.
\newblock Phase retrieval and saddle-point optimization.
\newblock {\em Journal of the Optical Society of America A}, 24(10):3289--3296,
  2007.

\bibitem{marquet-CT}
F.~Marquet, M.~Pernot, J.~F. Aubry, G.~Montaldo, L.~Marsac, M.~Tanter, and
  M.~Fink.
\newblock Non-invasive transcranial ultrasound therapy based on a {3D CT} scan:
  protocol validation and in vitro results.
\newblock {\em Physics in Medicine and Biology}, 54(9):2597--2614, 2009.

\bibitem{marsac2012mr}
L.~Marsac, D.~Chauvet, B.~Larrat, M.~Pernot, B.~Robert, M.~Fink, A.-L. Boch,
  J.-F. Aubry, and M.~Tanter.
\newblock {MR}-guided adaptive focusing of therapeutic ultrasound beams in the
  human head.
\newblock {\em Medical Physics}, 39(2):1141--1149, 2012.

\bibitem{maier}
N.~McDannold and S.~E. Maier.
\newblock Magnetic resonance acoustic radiation force imaging.
\newblock {\em Medical Physics}, 35(8):3748--3758, 2008.

\bibitem{lassas-x-ray}
E.~Niemi, M.~Lassas, and S.~Siltanen.
\newblock Dynamic {X}-ray tomography with multiple sources.
\newblock In {\em 8th International Symposium on Image and Signal Processing
  and Analysis (ISPA)}, pages 618--621. IEEE, 2013.

\bibitem{nocedal}
J.~Nocedal and S.~J. Wright.
\newblock {\em {N}umerical {O}ptimization}.
\newblock Springer, New York, 2nd edition, 2006.

\bibitem{pennes1948analysis}
H.~H. Pennes.
\newblock Analysis of tissue and arterial blood temperatures in the resting
  human forearm.
\newblock {\em Journal of applied physiology}, 1(2):93--122, 1948.

\bibitem{pursiainen}
S.~Pursiainen and M.~Kaasalainen.
\newblock Sparse source travel-time tomography of a laboratory target: accuracy
  and robustness of anomaly detection.
\newblock {\em Inverse Problems}, 30(11):114016, 2014.

\bibitem{casella}
C.~Robert and G.~Casella.
\newblock {\em {Monte Carlo Statistical Methods}}.
\newblock {Springer Science and Business Media, New York}, 2013.

\bibitem{sapareto1984thermal}
S.~A. Sapareto and W.~C. Dewey.
\newblock Thermal dose determination in cancer therapy.
\newblock {\em International Journal of Radiation Oncology Biology Physics},
  10(6):787--800, 1984.

\bibitem{shechtman2015phase}
Y.~Shechtman, Y.~C. Eldar, O.~Cohen, H.~N. Chapman, J.~Miao, and M.~Segev.
\newblock Phase retrieval with application to optical imaging: a contemporary
  overview.
\newblock {\em IEEE Signal Processing Magazine}, 32(3):87--109, 2015.

\bibitem{stewart-clinical}
E.~A. Stewart, J.~Rabinovici, C.~M.~C. Tempany, Y.~Inbar, L.~Regan, B.~Gastout,
  G.~Hesley, H.~S. Kim, S.~Hengst, and W.~M. Gedroye.
\newblock Clinical outcomes of focused ultrasound surgery for the treatment of
  uterine fibroids.
\newblock {\em Fertility and Sterility}, 85(1):22--29, 2006.

\bibitem{stuart-acta-numerica}
A.~M. Stuart.
\newblock Inverse problems: a {B}ayesian perspective.
\newblock {\em Acta Numerica}, 19:451--559, 2010.

\bibitem{tarantola}
A.~Tarantola.
\newblock {\em {I}nverse {P}roblem {T}heory and {M}ethods for {M}odel
  {P}arameter {E}stimation}.
\newblock {SIAM}, Philadelphia, 2005.

\bibitem{Haar-HIFU}
G.~{ter Haar} and C.~Coussios.
\newblock High intensity focused ultrasound: physical principles and devices.
\newblock {\em International Journal of Hyperthermia}, 23(2):89--104, 2007.

\bibitem{vogel}
C.~R. Vogel.
\newblock {\em {C}omputational {M}ethods for {I}nverse {P}roblems}.
\newblock {SIAM}, Philadelphia, 2002.

\bibitem{wu2004}
F.~Wu, Z.-B. Wang, W.-Z. Chen, W.~Wang, Y.~Gui, M.~Zhang, G.~Zheng, Y.~Zhou,
  G.~Xu, M.~Li, et~al.
\newblock Extracorporeal high intensity focused ultrasound ablation in the
  treatment of 1038 patients with solid carcinomas in {China}: an overview.
\newblock {\em Ultrasonics Sonochemistry}, 11(3):149--154, 2004.

\bibitem{zaporzan2013matmri}
B.~Zaporzan, A.~C. Waspe, T.~Looi, C.~Mougenot, A.~Partanen, and S.~Pichardo.
\newblock {MatMRI} and {MatHIFU}: software toolboxes for real-time monitoring
  and control of {MR}-guided {HIFU}.
\newblock {\em Journal of Therapeutic Ultrasound}, 1(1):1--12, 2013.

\end{thebibliography}

\end{document}